\documentclass{ieeeaccess}
\usepackage{cite}
\usepackage[numbers,sort&compress]{natbib}
\usepackage{amsmath,amssymb,amsfonts}
\usepackage{algorithm}
\usepackage{algorithmicx}
\usepackage{bm}
\usepackage{booktabs}
\usepackage{multicol}

\usepackage{graphicx}
\usepackage{textcomp}
\usepackage{float}
\usepackage{stfloats}
\usepackage{tabularx}
\def\BibTeX{{\rm B\kern-.05em{\sc i\kern-.025em b}\kern-.08em
    T\kern-.1667em\lower.7ex\hbox{E}\kern-.125emX}}

\begin{document}
\history{Date of publication xxxx 00, 0000, date of current version xxxx 00, 0000.}
\doi{10.1109/ACCESS.2022.DOI}

\title{Damage Maximization for Combat Network with Limited Costs}
\author{\uppercase{Jintao Yu}\authorrefmark{1}, \uppercase{Bing Xiao\authorrefmark{1}, and Yuzhu CUI}\authorrefmark{2}}
\address[1]{Air Force Early Warning Academy, Wuhan, 430019, China (e-mail:xb\_sky@126.com)}
\address[2]{Tsung-Dao Lee Institute, Shanghai Jiao Tong University, Shanghai, 210022, China (e-mail:yuzhu\_cui77@163.com)}

\tfootnote{This work was supported in part by National Nature Science Foundation under Grant No.~61502522, National University of Defense and Technology Research Program under Grant No.~JS20-10 and National Defense Science and Technology Innovation Foundation under Grant No.~GFCX425311.}

\markboth
{J. Yu, \headeretal: Damage Maximization for Combat Network with Limited Costs}
{J. Yu, \headeretal: Damage Maximization for Combat Network with Limited Costs}
%\headeretal 表示作者很多的
\corresp{Corresponding author: Jintao Yu (haleine@ustc.edu)}

\begin{abstract}
%Maximizing the damage of combat network plays a vital role in identifying the important nodes in combat system-of-system~(SOS). In order to protect or destroy the critical part of combat network more efficiently with less costs,
{\color{black} Maximizing the damage by attacking specific nodes of the combat network can efficiently disrupt enemies' defense capability, protect our critical units, and enhance the resistance to the destruction of system-of-system~(SOS). However, the modeling of the combat network damage is not practical enough. In this paper, we report a more realistic model to study the combat network damage maximization problems. By analyzing realistic situations, the cost of damage is redefined based on the network topology and the functional characteristics of nodes. The damage effect is also updated according to the combat network topology and operational capability. Hence, a cost-limited damage maximization model for the combat network is constructed. In addition, to obtain optimal solutions, an improved genetic algorithm~(IPGA) based on prior information is proposed. As a result, our method has a significant advantage in the feasibility and effectiveness compared with other algorithms in experiments. The attack pattern of the combat network and the convergence and complexity of the proposed algorithm are further explored. The improved model and algorithm, as well as the mined attack patterns, can provide support for military decisions. }
\end{abstract}


\titlepgskip=-15pt

\maketitle

\section{Introduction}
{\color{black}With the fast development of the social economy, there is an increasing number of complex networks significantly affecting our lives, such as the Internet, Internet of Things, transportation networks, power networks, communication networks and so on~\cite{b1}.}
{\color{black} These complex networks have stimulated the interest of many researchers. For instance, Barabási et al. explored the emergence of scaling in random networks and concluded that the nodes degree follows a scale-free power-law distribution~\cite{b29}, Bollobás et al.~\cite{b1_1} also proved the same results; Boccalettia et al.~\cite{b1_2} discussed the structure and dynamics of complex networks and Orsini et al.~\cite{b1_3} further investigated the randomness in real networks. The network has brought so many things together so it is an era of network interconnection now.}
However, while the network brings us considerable convenience, it also produces a lot of hidden dangers, {\color{black}such as computer viruses, epidemics and rumors,} which have led to plenty of disasters. How to protect the critical elements of these networks from attacks and maintain the robustness of those systems are the top questions to be answered~\cite{b2}. In addition, some harmful networks, such as terrorist relationship networks and drug trading networks, can be inversely traced or destroyed by searching {\color{black}a set of important nodes~\cite{b3}.}
From the perspective of network damage, mining the important nodes set can effectively protect or destroy the network. {\color{black}This idea is also applicable to combat networks.} Under the condition of informatization, there are complicated interactions among the various subsystems and weapon equipment units, forming a complex and large-scale combat network. In order to ensure the reliability and survivability of the combat system in a complex electromagnetic environment, it is essential to protect the combat network by finding the critical nodes~\cite{b3.1}. For the enemy, destroying important combat units can quickly disrupt their combat network, which allows us to gain an operational advantage. Hence, analyzing the disintegration strategy of the combat network can give strategic support for military actions against the adversary and guidance for improving our combat networks' invulnerability~\cite{b3.2}.  {\color{black}What we need to do is to find the best way to attack and destroy a network by node removal~\cite{b3.3}.}
In this paper, we will explore the important parts of a combat network from a view of the network attack, which enables us to protect our important facilities in a targeted manner and quickly disintegrate the {\color{black} enemies'} combat capabilities.

{\color{black}After extensive research on the robustness of the network in the past,} research topics related to network damage have also received considerable attention in recent years. The network damage maximization problem can be analogized to the influence maximization problem~\cite{b4}. The influence maximization problem aims to affect others as much as possible by a small set of specified seeds according to a certain propagation model~\cite{b5}. Similarly, the performance of the network is expected to be degraded as much as possible by destroying a limited set of critical nodes in the network damage maximization problem. However, conducting network prioritized attacks through the most common importance indicators, such as node degree centrality~\cite{b6,b7}, betweenness centrality~\cite{b8}, eigenvector centrality~\cite{b8}, closeness centrality~\cite{b9} and topological potential~\cite{b10}, often fails to achieve the optimal damage effect, despite their relatively low computational complexity. In addition, network damage maximization is as NP-hard as influence maximization. And most of the current methods for solving these combinatorial optimization problems require support from heuristic algorithms~\cite{b11}, swarm intelligence algorithms~\cite{b12}, or other algorithms to approximate the optimal solution. For example, Liu et al. used the combination of degree and betweenness as a heuristic factor to improve the search efficiency and proposed a greedy algorithm based on node importance to find an important set of nodes~\cite{b13}; Deng et al. utilized a tabu search optimization algorithm to find the optimal attack strategy for complex network disintegration~\cite{b14}; Considering the cascading failure in network damage, Zhang et al. tested a multi-objective optimization algorithm to maximize network damage in a simultaneous attack mode by adding the minimum number of attack nodes into the target~\cite{b15}. However, there are two disadvantages in the above studies, {\color{black} namely} ignoring the heterogeneity of damaged nodes and the importance of damage cost, which make the damage cost of each node almost at the same expensive level.

As for the network damage problem with heterogeneous nodes, Qi et al. studied the multiple links connected by layer nodes in multiplex networks, and reported an optimal disintegration strategy based on tabu search~\cite{b16,b17}; Deng et al. constructed the heterogeneous costs of network disintegration based on the network topology and discussed the disintegration strategy for complex networks under cost constraints~\cite{b18}; To reduce the total damage costs, Li et al.~\cite{b19} explored an elitism-based multi-objective evolutionary algorithm to investigate the robustness of complex networks based on Deng et al.'s research~\cite{b18}. The node heterogeneity and damage costs considered in these studies are mainly derived from the topology of complex networks. {\color{black}However, the functional characteristics of the network have not been taken into account,} which means the damage effect has less practical meaning. Hence, they can not be directly applied to combat networks. {\color{black} After the above analysis, a comparison} among the existing works are summarized in Table~\ref{tab_advantage}.

\begin{table*}[!htbp]
\newcolumntype{L}[1]{>{\raggedright\let\newline\\\arraybackslash\hspace{0pt}}m{#1}}
\newcolumntype{C}[1]{>{\centering\let\newline\\\arraybackslash\hspace{0pt}}m{#1}}
\newcolumntype{R}[1]{>{\raggedleft\let\newline\\\arraybackslash\hspace{0pt}}m{#1}}
\caption{The advantages and disadvantages of various methods. \label{tab_advantage}}
\centering
%\begin{minipage}[ht]{1\textwidth}
\begin{tabular}{L{4cm}L{6cm}L{6cm}}%{1.0\textwidth}{@{\extracolsep{\fill}}lll}%{p{4cm}p{6cm}p{6cm}}
  \toprule
  Method  & Advantages  & {\color{black}Disadvantages}\\
  \midrule
  Based on importance indicators~\cite{b6,b7,b8,b9,b10,b11,b12} & Relatively low computing complexity & Not necessarily the optimal solution, not including cost and heterogeneity	\\ \hline
  Heuristic algorithm~\cite{b13}  & Approximate optimal solution & Not including cost and heterogeneity	\\ \hline
  Tabu search optimization algorithm~\cite{b14}&Approximate optimal solution & Not including cost and heterogeneity\\ \hline
  Multi-objective optimization algorithm~\cite{b15} & Approximate optimal solution, including optimization goal of attack strength & Not including heterogeneity and cost constraints\\ \hline
  Tabu search optimization algorithm for multiplex networks~\cite{b16,b17}& Approximate optimal solution, including  heterogeneity& Not including cost \\ \hline
  Complex network disintegration algorithm~\cite{b19,b20}& Approximate optimal solution, including cost and heterogeneity  & Existing errors with the realistic model, not suitable for combat networks	\\
  \bottomrule
\end{tabular}

%\end{minipage}
\end{table*}

To improve the network damage maximization model by measuring the node heterogeneity and damage cost properly, and {\color{black}make it applicable to combat networks, it is necessary to consider} the damage maximization problem based on more realistic assumptions, especially for the combat network. {\color{black}The first step is to build a reasonable model of the combat network.} Li et al.~\cite{b20} constructed a heterogeneous network model of the combat system-of-system~(SOS) with limited information to disintegrate the network. According to the modeling method reported in~\cite{b20}, we establish a more realistic combat network model compared with previous studies. The second step is to estimate the damage cost. Since there are limited resources that can be allocated in the practical battle, the damage capability is limited by a certain cost. {\color{black}As mentioned in the preceding,} the damage cost can not be estimated simply based on the network topology. In order to reassess the damage cost more precisely, it is necessary to introduce the node capability and defense efficiency into the calculation model. What is more important, different levels of strike intensity according to the combat determination should also be considered in combat network damage. In this paper, we establish a damage maximization model which can attack a set of certain nodes in a combat network with limited costs. We also propose an optimal algorithm to approximate the best damage result. In addition, the performance of this algorithm is examined under various conditions and {\color{black}constraints.}

This paper is organized as follows. In Section~\ref{sec:2}, the combat network model is established at first. Then a realistic calculation method of {\color{black} the} combat network damage cost {\color{black}as well as the} damage effect is proposed, respectively. The damage maximization model for a combat network with limited costs is constructed accordingly. To solve the mathematical optimization problem, Section~\ref{sec:3} presents an improved genetic algorithm~(IPGA) based on {\color{black}the} prior information. In Section~\ref{sec:4}, the comparative experiments based on the simulated combat networks are carried out to illustrate the {\color{black}feasibility and effectiveness} of the proposed algorithm and its superiority over common algorithms. The attack law and algorithmic properties are further explored. {\color{black}Finally, the conclusions, {\color{black} limitations} and future directions are remarked on in Section~\ref{sec:5}.}

%%%%%%%%%%%%%%%%%%%%%%%%%%%%%%%%%%%%%%%%%%
\section{Damage Maximization Model of Combat Network with Limited Costs}
\label{sec:2}
\subsection{Combat Network Model}
{\color{black}In general, when we apply network science to the study of the modeling problem of combat SOS, combat units and the interactions} among them are often represented as nodes ($v_{\rm i} \in V$) and edges ($e_{\rm i} \in E$) of an undirected complex network, denoted as $G(V, E)$. The scales of nodes and edges are $N=|V|$ and $M=|E|$, respectively. Note that different combat units may have different functions in combat SOS, which means that one unit can either be regarded as one node by itself or be decomposed into several nodes with different sub-functions. In this paper, we will adopt the early warning intelligence combat SOS {\color{black}as the specific research context.} Accordingly, the nodes can be divided into
four categories by their functions, namely the intelligence obtaining node O, intelligence processing node P, commanding and decision node D, and attack/damage node A. In addition, there are different relations among them, such as communication, intelligence analysis and transmission, cooperative detection, coordinate command, and fire strike. For the early warning intelligence combat network, the specific types of node connections mainly include O--O, O--P, P--P, P--D, D--D, and D--A, which constitute a complex network. To {\color{black}express} the heterogeneous information contained in the actual combat network, we consider both the topology and the functional attributes when measuring the damage costs.

\begin{table}[!htbp]
\newcolumntype{L}[1]{>{\raggedright\let\newline\\\arraybackslash\hspace{0pt}}m{#1}}
\newcolumntype{C}[1]{>{\centering\let\newline\\\arraybackslash\hspace{0pt}}m{#1}}
\newcolumntype{R}[1]{>{\raggedleft\let\newline\\\arraybackslash\hspace{0pt}}m{#1}}
\caption{List of symbols. \label{tab_glossary}}
\centering
\resizebox{0.5\textwidth}{!}{
\begin{tabular}{C{1.5cm}C{6cm}}%{p{2.5cm}p{5cm}}
  \toprule
  Symbol  & Meaning	\\
  \midrule
  $G$	    & The complex network		\\  \hline
  $V$	    & The node set of the complex network		\\  \hline
  $v_i$   & The node in the complex network   \\  \hline
  $N$   	& The node number of the set of nodes   \\  \hline
  $E$	    & The edge set of the complex network		\\  \hline
  $e_i$   & The edge in the complex network  \\  \hline
  $M$   	& The {\color{black}edge} number of the set of edges \\  \hline
  $c_i(\bm{C})$     & The damage cost of node $v_i$, and the vector of them is $\bm{C}$	\\  \hline
  $C_\rm{max}$      & The maximum damage cost\\ \hline
  $d_i$	          & The degree of node $v_i$	\\  \hline
  $\lambda$	      & The correction coefficient of the damage cost	\\  \hline
  $\gamma$	      & The power parameter reflecting the sensitivity to damage costs	\\  \hline
  $\bm{S}$		  & The adjacency matrix of the combat network\\  \hline
  $\tilde{\bm{S}}$  & The accessibility matrix of the entire combat network obtained from $\bm{S}$	\\  \hline
  $\bm {X}$	      & The encoded vector of a set of nodes, it is also the symbol of a chromosome denoted as $\bm{X}^\rm{i}_\rm{p}$\\  \hline
  $\hat{\bm{X}}_\rm{p}$      	& The population of  chromosome	\\  \hline
  $\hat{\bm{Y}}_\rm{p}$	    & The population of  chromosome after crossover procedure	\\  \hline
  $\hat{\bm{Y}}'_\rm{p}$		& The population of  chromosome after mutation procedure	\\  \hline
  $\bm{H}$	    & The index vector of different complex network centralities\\  \hline
  $S_\rm{links}$  & The amount of total intelligence effectiveness links	\\  \hline
  $S_\rm{huge}$   & The largest component scale of the combat network		\\  \hline
  $\alpha$	    & The preference parameter to metric $S_\rm{Links}$ and $S_\rm{Huge}$ 		\\  \hline
  $R$	            & The measure of damage {\color{black}effect} of the combat network		\\  \hline
  $cn_1, cn_2$	& The amount of chromosome crossover or mutation point\\  \hline
  $P_\rm{c}$	    & The crossover probability		\\  \hline
  $P_\rm{m}$	    & The mutation probability 		\\  \hline
  $N_\rm{p}$	    & The {\color{black}population} size		\\  \hline
  $gen$	        & The upper limit of iteration times		\\  \hline
  $L$	            & The strength of attack 		\\  \hline
  $p$             & The iteration times of the IPGA			\\
  \bottomrule
\end{tabular}
}
\end{table}

For the convenience of reading, a glossary of the notations used in this paper is presented in Table~\ref{tab_glossary}.

\subsection{Calculation of Damage Costs}
For a complex network, Deng et al. adopted the exponential power of node degree as the damage cost of each node~\cite{b18}, which can measure the damage costs well. We will improve the calculation of damage costs based on this. %However, for a combat network with heterogeneous nodes, it is insufficient to measure the damage cost only according to the network topology.

In the practical combat network, different nodes possess different levels of importance according to their functional attributes~(such as communication capabilities) and self-defense attributes~(such as deployed location, maneuverability, camouflage capability, defense capability, and self-repair capability), which will affect the damage costs significantly. It is natural to apply a method of multi-attribute evaluation to study the importance of different nodes. Then the damage costs can be adjusted based on node importance.
%, which can be used to improve the damage costs calculation rather than only based on topology.
Taking the intelligence obtaining node O as the reference, we define its correction coefficient $\lambda$ as 1 and the correction coefficients of the remaining nodes P, D, and A as 1.4, 1.6, and 1.1, respectively. We denote the corrected node damage cost vector as $\bm{C}=[c_1,c_2,\cdots ,c_{\rm N}]$, then the damage cost of each node is expressed as

\begin{equation}
    c_{\rm i}=\lambda {d_{\rm i}}^{\gamma}\,,
\end{equation}
where $d_{\rm i}$ is the degree of node $v_{\rm i}$, $\gamma$ is a power parameter reflecting the sensitivity to damage costs and it is all the same for nodes in the combat network~\cite{b18}. When $\lambda$ and $\gamma$ {\color{black}are} both equal to 1, all network nodes will have the same damage costs.

During the damage process of a combat network through finite nodes, the nodes with high damage costs will consume a lot of resources. Therefore, it is necessary to select some suitable targets to maximize the damage effect when there is a cost upper bound constraint. The cost upper bound is generally measured by a certain percentage of {\color{black}the} total cost, represented as

\begin{equation}
    C_{\max}=\rho \sum\limits_{i = 1}^N {{c_{\rm i}}}\,,
\end{equation}
where $\rho \in [0,\,1]$ is the proportional parameter of the total cost constraints.

\subsection{Measure of Damage Effect}
In the normal network damage problem with a damage cost constraint, the measure of damage effect is usually represented by the network efficiency~(common indicator for communication networks) or the relative size of {\color{black}the maximum connected patch.}
%But for the combat network damage problem with cost constraints, not only the nodes are heterogeneous, but also the combat network has obvious physical meaning in actual combat activities. It is definitely inappropriate to still use the above-mentioned indicator as a measure of damage effect.
{\color{black}Considering the structural characteristics and the actual operational meaning of a combat network itself, the damage effect of a combat network is supposed to be described from two aspects.}

The first aspect is about the damage effect measure according to {\color{black}the topology of a combat network.} The scale of the maximal component is selected as an index of the damage effect of a combat network. One individual component is a subgraph with connectivity and isolation in the network. As a result, the maximal component is a subgraph with the largest node scale in the network, which is denoted as $S_{\rm{huge}}$. With the increase of the scale of the maximal component in the combat network, {\color{black}the interconnections among combat network nodes get closer and the information flow efficiency of the combat network becomes higher.} In this paper, we adopt the approach of node shrinking to iteratively calculate the scale of the maximal component~\cite{b21}.

\begin{figure}
\centering
\includegraphics[width=6.5cm]{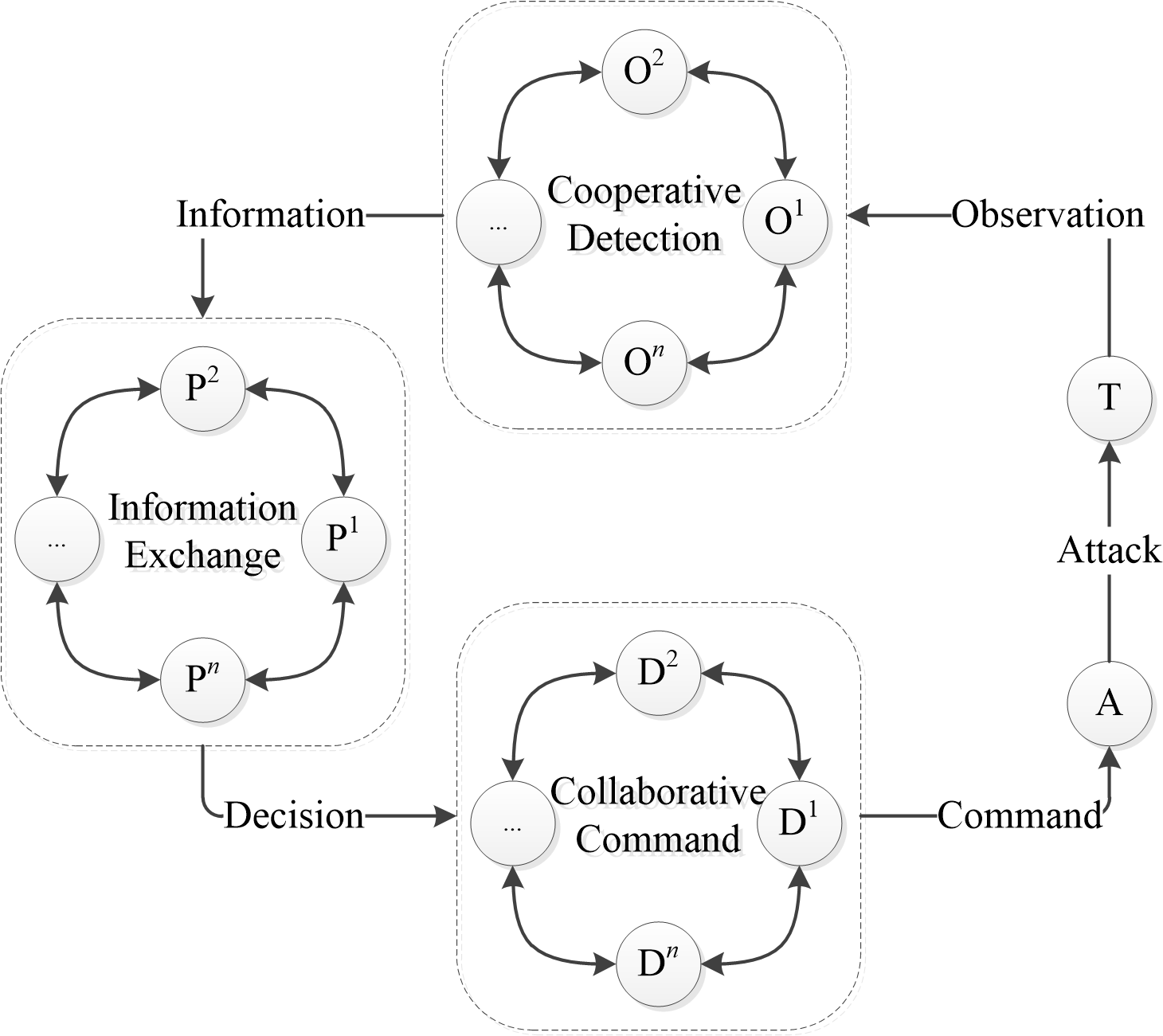}
\caption {Generalized IELP diagram.}\label{fig_loop}
\end{figure}

The second aspect is about the damage effect measure based on the operational capability of a combat network. The number of attack links can be used to describe the operational capability of a combat network~\cite{b22}. {\color{black}In the case of the combat network for early warning intelligence SOS,} the number of intelligence effectiveness links~(IELKs) $S_{\rm{links}}$ is introduced to measure the operational capability. According to the idea of Boyd's OODA cycle~\cite{b23}, the combat network exerts operational capability by forming an IELK of `intelligence obtaining--intelligence processing-commanding and decision--attack and damage' around the target,
namely the OODA attack link. For a more general IELK, the mutual coordination among the intelligence obtaining node O, the intelligence processing node P, and the commanding and decision node D should also be taken into consideration. The flow of generalizing IELK with a target is demonstrated in a generalized intelligence {\color{black}effectiveness} loop~(IELP), which is shown in Figure~\ref{fig_loop}. Since the node cooperation in generalized IELK may lead to an infinite long link, seven types of IELKs commonly used in practice are selected as the basis for quantity calculation~\cite{b22}. The detailed description of those IELKs is indicated in Table~\ref{tab_links}.
%to test different conditions. The detailed description of tested IELKs is indicated in Table~\ref{tab_links}.

\begin{table}[!htbp]
\newcolumntype{L}[1]{>{\raggedright\let\newline\\\arraybackslash\hspace{0pt}}m{#1}}
\newcolumntype{C}[1]{>{\centering\let\newline\\\arraybackslash\hspace{0pt}}m{#1}}
\newcolumntype{R}[1]{>{\raggedleft\let\newline\\\arraybackslash\hspace{0pt}}m{#1}}
\caption{Seven types of common IELKs and their definitions. \label{tab_links}}
\centering
\begin{tabular}{L{2.3cm}L{5.2cm}}
  \toprule
  IELK   & Definition	\\
  \midrule
  %\hline
  O--P--D--A	    & The standard IELK		\\   \hline
  O--O--P--D--A   	& IELK with cooperative detection	\\  \hline
  O--P--P--D--A	    & IELK with information interaction	\\  \hline
  O--P--D--D--A		& IELK with coordinated command		\\  \hline
  O--O--P--P--D--A	& IELK with cooperative detection and information interaction	\\  \hline
  O--O--P--D--D--A	& IELK with cooperative detection and coordinated command		\\  \hline
  O--O--P--P--D--D--A	& IELK with cooperative detection, coordinated command, and information interaction			\\
  \bottomrule%\hline
\end{tabular}
\end{table}

The accessibility matrix $\tilde{\bm{S}}$ of the entire combat network is calculated according to the adjacency matrix $\bm{S}$ of the combat network:
\begin{equation}\label{keda_matrix}
\begin{array}{l}
{(\bm{S} + \bm{I})^{(1)}} \ne {(\bm{S} + \bm{I})^{(2)}} \ne  \cdots \\
 \qquad  \qquad \,\,\ne {(\bm{S} + \bm{I})^{(r)}} = {(\bm{S} + \bm{I})^{(r + 1)}} = \tilde {\bm{S}}\,,
\end{array}
   % {(S + I)^{(1)}} \ne {(S + I)^{(2)}} \ne  \cdots  \\
    %\ne {(S + I)^{(r)}} = {(S + I)^{(r + 1)}} = \tilde{S}
\end{equation}
where $I$ is the identity matrix. Equation~(\ref{keda_matrix}) is the power boolean operations on $(\bm{S} + \bm{I})$, {\color{black}and $r+1$ is the times of power multiplication. The connectivity among intelligence obtaining nodes and attack/damage nodes can be easily obtained through the accessibility matrix. For any node $O_{\rm i}$ and node $A_{\rm j}$, if {\color {black}$\bm{\tilde S}(i,j)=1$,} then $O_{\rm i}$ can reach $A_{\rm j}$ according to a path with practical meaning. Let $\bm{S}(j,i)=1$,} we can get

\begin{equation*}
  \bm{S} = \left[ {\begin{array}{*{20}{c}}
  {{\bm{S}_{{\rm{OO}}}}}&{{\bm{S}_{{\rm{OP}}}}}&0&0\\
  0&{{\bm{S}_{{\rm{PP}}}}}&{{\bm{S}_{{\rm{PD}}}}}&0\\
  0&0&{{\bm{S}_{{\rm{DD}}}}}&{{\bm{S}_{{\rm{DA}}}}}\\
  {{\bm{S}_{{\rm{AO}}}}}&0&0&0
\end{array}} \right],
\end{equation*}
then the number of IELPs is the trace of the product of corresponding nodes' accessibility matrices, which is also the number of IELKs. Taking the `O--P--D--A' link as an example, the number of this link can be calculated by
\begin{equation}
    {S_{\rm {link}}} = tr(S_{\rm {OPDAO}}) = tr({\bm{S}_{\rm {OP}}} \times {\bm{S}_{\rm {PD}}} \times {\bm{S}_{\rm {DA}}} \times {\bm{S}_{\rm {AO}}})\,.
\end{equation}

{\color{black}This calculation process for the remaining types of links is the same. %\textcolor[rgb]{0.00,1.00,0.00}{Then we can get the amount of total links is}
{\color{black} Then we can obtain the number of IELKs of all 7 types as}  $S_{\rm {links}}=\sum\nolimits_{\rm i = 1}^7 {S_{\rm {link}}^{\rm i}} $.

After two aspects of analysis, we can calculate the measure of the damage effect. The damage effect of a combat network in this paper should be calculated by relative metrics.} For the original combat network $G$ which has not been attacked, the initial largest component scale is $S_{\rm {huge}}(G)$ and the amount of IELKs is $S_{\rm {links}}(G)$. For the attacked combat network $G'$, the corresponding largest component scale and the amount of IELKs are $S_{\rm {huge}}(G')$ and $S_{\rm {links}}(G')$, respectively. We can measure the damage effect of a combat network in the following expression:
\begin{equation}\label{eq_R}
    R = 1 - \left[ {\alpha \frac{{{S_{\rm {link}}}(G')}}{{{S_{\rm {link}}}(G)}} + (1 - \alpha )\frac{{{S_{\rm {huge}}}(G')}}{{{S_{\rm {huge}}}(G)}}} \right]\,,
\end{equation}
where $\alpha$ is a proportion parameter indicating the preference for two different metrics. The default value of $\alpha$ is 0.5.

\subsection{Damage Maximization Model of Combat Network Under Cost Constraints}
The damage maximization model of combat network under cost constraints can be described as a combinatorial optimization problem of selecting $L$ nodes from $N$ nodes to damage under the given damage cost constraints $C_{\max}$, where $L$ is the damage intensity. In order to avoid complex operations on the adjacency matrix of the combat network directly %\textcolor[rgb]{0.00,1.00,0.00}{and describe the subsequent algorithm more conveniently, the nodes are first encoded to calculate the damage effect on the combat network after removal of the nodes, and in this mode, the description of mathematical optimization can be carried out smoothly.}
{\color{black} and to describe the following optimization algorithm more conveniently, the nodes are first encoded to identify the nodes being removed, then the damage effect on the combat network after node removal can be calculated. In this mode, the description of mathematical optimization can be carried out smoothly.}
{\color{black}After obtaining the optimal encoding vector of the programming model, the decoding algorithm is used to map the result back to the nodes in the combat network, and then the set of nodes that need to be destroyed in the combat network is clear.} The specific steps of the encoding algorithm and decoding algorithm are shown in the {\color{black}Algorithm}~\ref{tab_coding} and~\ref{tab_decoding}, respectively.

\begin{algorithm}[!htbp]
	\caption{$\bm{X}=Encoding(G, Q)$ \label{tab_coding}} %算法的名字
	\hspace*{0.02in} {\bf Input:} %算法的输入， \hspace*{0.02in}用来控制位置，同时利用 \\ 进行换行
	Combat network $G(V, E)$ and attacked nodes set $Q$	\\
	\hspace*{0.02in} {\bf Output:} %算法的结果输出
	Encoded vector $\bm{X}=[x_1,\,x_2,\,\cdots,\,x_{\rm N}]	$
	\begin{algorithmic}[1]
		\State \,Initialize $\bm{X} = [x_1,\,x_2,\,\cdots,\,x_{\rm N}] = [0,\,0,\,\cdots ,\,0]$	
		\State\,\textbf{for} $i=1:N$
		\State\qquad \textbf{if} $v_{\rm i} \in Q$
		\State\qquad \qquad $x_{\rm i}=1$
		\State\qquad \textbf{end if}
		\State\,\textbf{end for}		
	\end{algorithmic}
\end{algorithm}
\begin{algorithm}[!htbp]
	\caption{$Q=Decoding(G,\bm{X})$ \label{tab_decoding}} %算法的名字
	\hspace*{0.02in} {\bf Input:} %算法的输入， \hspace*{0.02in}用来控制位置，同时利用 \\ 进行换行
	Combat network $G(V, E)$ and encoded vector $\bm{X}=[x_1,\,x_2,\,\cdots,\,x_{\rm N}]$\\
	\hspace*{0.02in} {\bf Output:} %算法的结果输出
	Attacked nodes set $Q$
	\begin{algorithmic}[1]
		\State \,Initialize $Q = \emptyset $	
        \State \,\textbf{for} $i=1:N$
        \State \qquad \textbf{if} $x_{\rm i} =1 $
        \State \qquad \qquad $Q = Q \cup \{ v_{\rm i} \}$
        \State \qquad \textbf{end if}	
        \State \,\textbf{end for}		
	\end{algorithmic}
\end{algorithm}

Based on the encoded vector $\bm{X}$ obtained by the encoding algorithm $Encoding(G, Q)$, the damage measure of Equation~(\ref{eq_R}) can be expressed as
\begin{equation}
    \Phi (\bm{X} = [{x_1},{x_2}, \cdots ,{x_{\rm N}}]) = R(Encoding(G,Q))\,.
\end{equation}
Hence, the damage maximization model for the combat network by attacking {\color{black}a fixed number of nodes with limited costs is as follows:}
\begin{equation}\label{eq_model}
    \begin{array}{l}
\max {\rm{ }}\Phi (\bm{X} = [{x_1},\,{x_2}, \,\cdots ,\,{x_{\rm N}}])\\
s.t.\left\{ \begin{array}{l}
{\bm{C}^T}\bm{X} \le {C_{\max }}\\
\sum\nolimits_{i = 1}^N {{x_{\rm i}}}  = L\\
{x_{\rm i}} = 0 \, or \, 1, \, i = 1, \, 2, \, \cdots,\,N\,.
\end{array} \right.
\end{array}
\end{equation}

The Equation~(\ref{eq_model}) is a zero-one integer programming model. {\color{black}Let's consider a simple example: If the nodes scale of the combat network is 100 and 10 nodes are chosen to be attacked, then the solution space size of the programming model will reach $C_{100}^{10} \approx 1.7 \times {10^{13}}$. In reality, the scale of the combat network is much larger than this, so the magnitude of the solution space will be larger. It is obviously inappropriate if we use traditional methods such as traversal algorithms, because they could consume a lot of computation and cause very high time complexity. In addition, since the objective function value needs to be calculated independently, it is not a simple explicit value. As a result, conventional combinatorial optimization algorithms such as the branch and bound method, can not work. For this kind of problem, it is a good choice to solve it by means of heuristic intelligent optimization algorithms. In the following, we will propose an improved intelligent optimization algorithm to approximate the optimal solution.}

%%%%%%%%%%%%%%%%%%%%%%%%%%%%%%%%%%%%%%%%%%
\section{Solution of Optimization Model Based on Improved Genetic Algorithm}
\label{sec:3}
\subsection{Classical Genetic Algorithm}
Genetic algorithm~(GA) forms an adaptive global optimization search mechanism by simulating the genetic and evolutionary processes of organisms in nature~\cite{b24}. It draws on the concepts of selection, inheritance, and variation in genetics to achieve the improvement of individual fitness in a population. The steps of the classical GA~\cite{b25} are generally:

\textbf{Step1:} Initialize the population;

\textbf{Step2:} Calculate individual fitness;

\textbf{Step3:} Perform roulette-wheel selection;

\textbf{Step4:} Determine whether the termination condition is reached. If the condition is satisfied, the algorithm ends; otherwise, perform crossover and mutation operations and then go to Step2.

\subsection{Improved Genetic Algorithm}
Due to the large scale of the solution space, there are still problems such as slow convergence speed and high computational complexity when the classical GA is applied to study the combat network damage problem. In addition, the cost constraints and the specified damage intensity require special processing when using GA. {\color{black}Hence, it is necessary to improve GA with the purpose of its application to practical problems.}

\subsubsection{Handling of Cost Constraints}
In the combinatorial optimization model of Equation~(\ref{eq_model}), it is required that the total cost of the nodes to be damaged does not exceed the given cost threshold. As for this constraint, the idea of penalty function is used to handle it equivalently. When calculating the individual fitness function value, the result of fitness will be penalized if the expected damage cost is greater than the given cost constraints.
That is, the fitness function value will be a relatively small negative number such as $-2$ while the ideal result is 1 if the expected damage cost is within the constraints.

\subsubsection{Initialization}
{\color{black}When the classical GA produces the first generation of population, all initial solutions are generated at random with equal probability.} On the one hand, the fitness value may not be satisfying, and on the other hand, it may lead to premature convergence. Therefore, initializing the population is a key step for this heuristic algorithm~\cite{b26}, and reasonable initialization can effectively improve the performance of the algorithm.

%In fact, it is suggested that the heterogeneity of combat network nodes can be used for initialization.
{\color{black}In fact, the combat network nodes vary in topology and functions, thus some prior knowledge can be obtained from them for initialization. When finding the set of important nodes by means of network damage, the node centrality can be introduced to adjust the initialization strategy.} For one thing, we hope to obtain an initial population with {\color{black}specific guidance through this prior information. For another, we want to maintain a good diversity of the population.} Therefore, the initialization is divided into two steps. In the first step, the nodes are sorted based on the degree centrality, betweenness centrality, and topological potential, respectively. Then each node will be assigned a probability according to the value of the above metrics. Then $L$ times sampling without replacement are performed with three different indicators, and $n_{\rm p}$ initial chromosomes are {\color{black}generated, respectively.} In the second step, the remaining $(N_{\rm p}-3n_{\rm p})$ chromosomes are obtained by randomly generated sequences, {\color{black}and the number of nodes selected for the chromosome should be $L$. In this way, {\color{black}these two steps} generate an initial population of size $N_{\rm p}$~($N_{\rm p}$ is an even number) in total,} which not only has specific information to help improve the convergence speed and optimization accuracy of the algorithm but also maintains the diversity during the population differentiation. As a result, the generated initial solutions can be more reasonable. {\color{black}Denote the initial population matrix as $\hat{\bm{X}}_0$, and each chromosome vector in the population as $\bm{X}^\rm{i}_0 (i=1,2,\cdots,N_{\rm p})$.}

\subsubsection{Crossover and Mutation}
Crossover is to randomly match an individual in the population and exchange some chromosomes between them according to a certain crossover probability. Generating new offspring by crossover can improve the {\color{black}search capability of GA.} In order to maintain the stable damage intensity, the crossover operation in the IPGA must be symmetric, that is, after chromosome A and chromosome B perform {\color{black}a} 0--1 exchange at one position, there must be a corresponding 1--0 exchange operation at another position at the same time. The specific steps of the crossover operation are:

\noindent\textbf{Step1:} Initialize $i=0$, $cn_1=0$, $cn_2=0$;\\
\textbf{Step2:} {\color{black}Let $i=i+1$, then choose an unselected chromosome pair and generate a random number $r$. If $r<P_{\rm c}$,} go to Step3, otherwise go to Step2;\\
\textbf{Step3:} Randomly select several positions that need to be crossed, denoted as $loc$. {\color{black}If the chromosome pair codes at the position $l_j \,(l_j \in loc)$ are 0 and 1, $cn_1=cn_1+1$. And if the chromosome pair codes at the position $l_j$ are 1 and 0, $cn_2=cn_2+1$. Then randomly select $k=\min (cn_1,cn_2)$ points to exchange value simultaneously as to the above two code types. After that, let $cn_1=0$ and $cn_2=0$.}\\
\textbf{Step4:} If $i=N_{\rm p} / 2$, the crossover operation is completed, otherwise go to Step2 until the termination condition is satisfied.

Similarly, when the mutation operation brings a new search space to the algorithm by reversing the code value according to the mutation probability, the principle of symmetric mutation should also be followed. The specific steps are:

\noindent\textbf{Step1:} {\color{black}Initialize $i=0$;}\\
\textbf{Step2:} {\color{black}Let $i=i+1$, then choose an unselected chromosome and generate a random number $r$. If $r<P_{\rm m}$,} go to Step3, otherwise go to Step2;\\
\textbf{Step3:} {\color{black}Randomly select some points that need to be mutated, denoted as $loc$, and the length of $loc$ should be less than $L$. If the chromosome code at the position $l_j \,(l_j \in loc)$ is 1, a mutation point $l_j'$ is also randomly selected at the position where the coded value is 0,} and the values at these positions are reversed;\\
\textbf{Step4:} If $i=N_{\rm p}$, the mutation update is completed, otherwise go to Step2.

\subsubsection{Procedure of Algorithm}
\begin{figure}[!htbp]
\centering
\includegraphics[width=7cm]{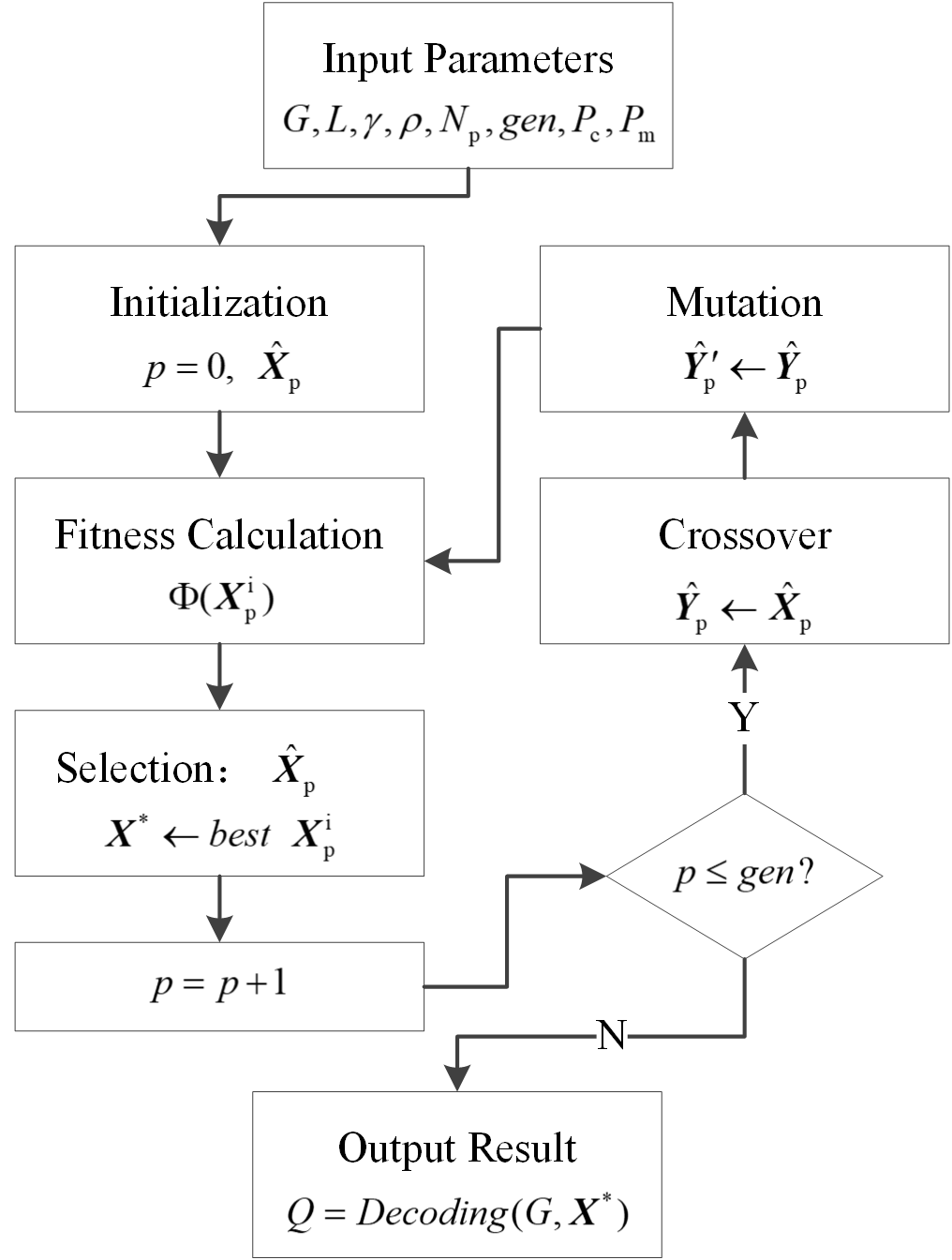}
\caption {\color{black}Flow diagram of the IPGA.}\label{fig_GA}
\end{figure}

On the basis of the above analysis, the algorithm flow chart of the IPGA to optimize the maximum damage effect of combat network {\color{black}with} limited costs is given, as shown in Figure~\ref{fig_GA}. The input parameters of the algorithm include the combat network model $G$, the damage intensity $L$, the cost power parameter $\gamma$, the cost constraint ratio $\rho$, and the initial parameters of GA, {\color{black}which are} the population size $N_{\rm p}$, the upper limit of iteration times $gen$, the crossover probability $P_{\rm c}$, and the mutation probability $P_{\rm m}$. The output of the algorithm is {\color{black}a set of the damaged nodes $Q$.} At the beginning, the population of the chromosome $\hat{\bm{X}}_\rm{p}$ is initialized when $p=0$. The Equation~(\ref{eq_Xp}) reflects the relations among the population and chromosomes.
\begin{equation}\label{eq_Xp}
    \hat{\bm{X}_\rm{p}} = \left[ \begin{array}{c}
    {\bm{X}^1_\rm{p}}\\
    {\bm{X}^2_\rm{p}}\\
     \cdots \\
    {\bm{X}^{N_{\rm{p}}}_\rm{p}}
    \end{array} \right] = \left[ \begin{array}{c}
    {[x_1^1,x_2^1, \cdots ,x_N^1]_\rm{p}}\\
    {[x_1^2,x_2^2, \cdots ,x_N^2]_\rm{p}}\\
     \cdots \\
    {[x_1^{{N_{\rm{p}}}},x_2^{{N_{\rm{p}}}}, \cdots ,x_N^{{N_{\rm{p}}}}]_\rm{p}}
    \end{array} \right].
\end{equation}

{\color{black}The best chromosome $\bm{X}^*$ in current generation is selected by calculating the fitness value of each chromosome $\bm{X}^i_\rm{p}$ in the population $\hat{\bm{X}_\rm{p}}$ at the $p_{th}$ generation.} If the termination condition is not met, the crossover and mutation procedure will be carried out, and the population $\hat{\bm{X}}_\rm{p}$ will be updated to $\hat{\bm{Y}}_\rm{p}$ and $\hat{\bm{Y}}'_\rm{p}$, respectively. {\color{black}And the set $Q$ is finally obtained by the decoding operation on the best chromosome $\bm{X}^*$. }

\section{Experiments and Analysis}
\label{sec:4}
In order to study the performance of the IPGA, some simulation experiments of combat network damage {\color{black}with cost constraints are carried out based on the model network.} The effectiveness and superiority of the IPGA are verified by comparing it with other algorithms based on the prioritization of centrality indicators. {\color{black}Then the optimal gap effect of IPGA and the attack law of maximizing the damage for the combat network under cost constraints are further explored.} Finally, the time complexity of the algorithm is demonstrated by theoretical analysis and simulation experiments.

\subsection{Construction of Combat Network}
{\color{black}In general, the combat network} consists of several subnets, including the intra-layer subnets formed by nodes of the same type and the inter-layer subnets formed by nodes of different {\color{black}types.}
%In a combat network, the subnets are formed by interconnections among most of the same type of nodes.
Due to the lack of real data and confidentiality, here we use different model networks to simulate the intra-layer subnets {\color{black}in the real world.} The ER random network~\cite{b27,b28}, BA scale-free network~\cite{b29} and Goh scale-free network~\cite{b30} are chosen to generate subnets. The ER random network {\color{black}adopts} a probability to connect two nodes randomly. {\color{black}The BA network tends to add nodes to those possessing a larger node degree, and the distribution of node degree of it obeys the power law whose parameter is about 3. The Goh network is also a scale-free network, and it is a network with an adjustable power parameter.} As for the connection between different subnets, nodes are connected randomly. And for different model networks, the size of the subnet is set as $N_{\rm O}=50$, $N_{\rm P}=40$, $N_{\rm D}=30$, $N_{\rm A}=30$ and the size $N$ of the total network is 150. The connection probability of the ER random network is $f_{\rm {OO}}=0.02$, $f_{\rm {PP}}=0.05$, $f_{\rm {DD}}=0.05$, $f_{\rm {AA}}=0.03$; the parameters of the BA scale-free network are $m_0=5$ and $m=3$; the parameters of the Goh scale-free network are $\beta =2.3$ and $\left\langle k \right\rangle = 6$, and the connection probability between subnets is uniformly set to 0.03. In order to reduce the randomness of the simulation experiments and improve the reliability of results, each type of model network is repeatedly generated 100 times, {\color{black}resulting in} a total of $3 \times 100$ networks.

\subsection{Simulation Experiments}
The simulation experiments based on the generated model network are divided into {\color{black}four parts: the damage effect comparison with different algorithms, the damage effect comparison with different parameters and network size, the optimal gap effect for reduced size networks, and the exploration of the attack law in combat network damage maximization problem based on the IPGA.} The simulation software is Matlab 2016b with Windows\,10, and the hardware configuration is Intel(R) Core(TM) i7-10750H CPU @ 2.60\,GHz.

\subsubsection{Comparison with Different Algorithms}
The comparison algorithms of the IPGA mainly include importance--first algorithms based on indicators such as degree centrality, betweenness centrality, eigenvector centrality, closeness centrality, and topology potential. {\color{black}For each centrality indicator, the larger the indicator value, the more important the node is.} In order to meet the cost constraints, a programming model is established when selecting important nodes to attack by different indicators. The goal is to maximize the sum of the {\color{black}indicator values} of the selected nodes, let the index vector be $\bm{H}$, then the nodes based on the importance indicator are selected according to the following model:
\begin{equation}\label{eq_HX}
  \begin{array}{l}
\qquad \quad \max {\bm{H}^T}\bm{X}\\
s.t.\left\{ \begin{array}{l}
{\bm{C}^T}\bm{X} \le {C_{\max }}\\
\sum\nolimits_{\rm i = 1}^{\rm N} {{x_{\rm i}}}  = L\\
{x_{\rm i}} = 0 \, or \, 1, \, i = 1,\,2,\, \cdots,\,N\,.
\end{array} \right.
\end{array}
\end{equation}

\begin{figure*}[!htbp]
%\begin{adjustwidth}{-\extralength}{0cm}
\centering
 \begin{minipage}{1\textwidth}
  \includegraphics[width=6cm]{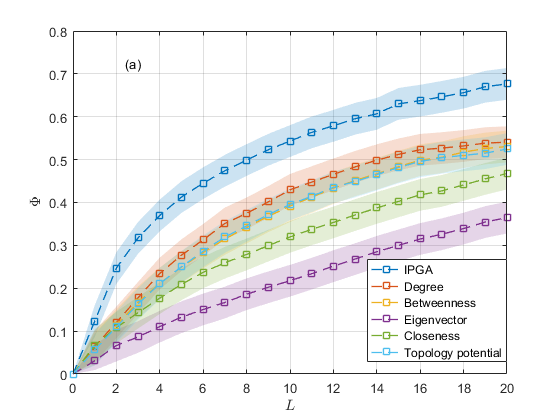}
  \includegraphics[width=6cm]{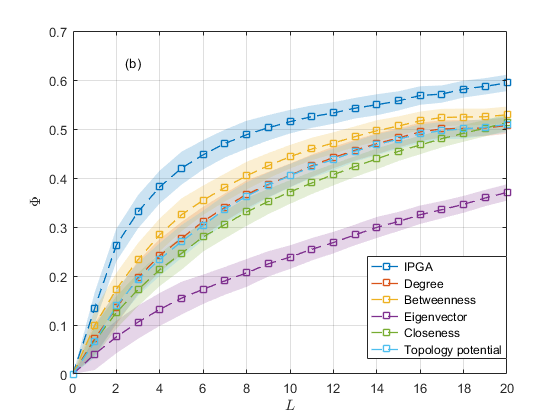}
  \includegraphics[width=6cm]{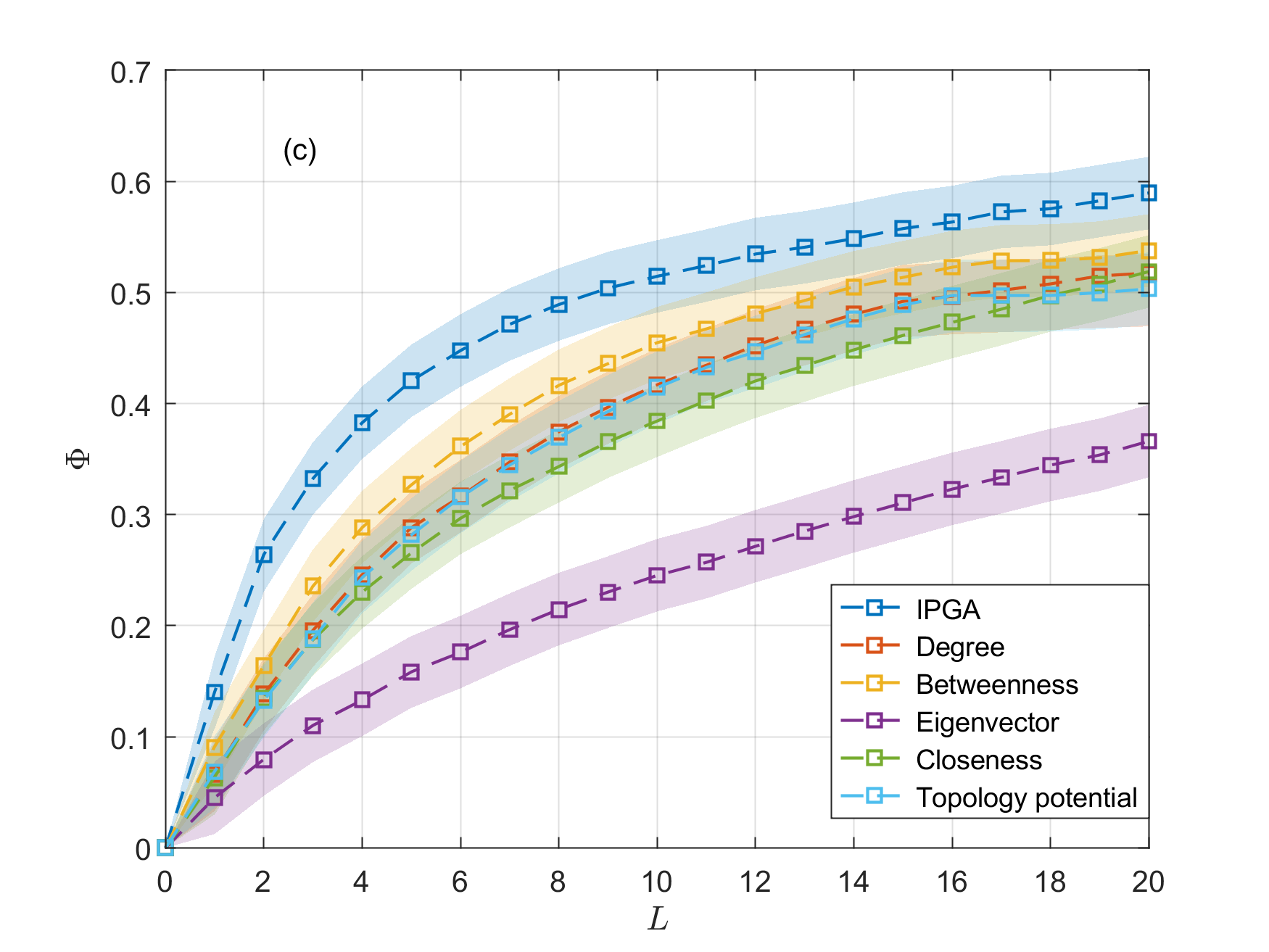}
  \end{minipage}
%\end{adjustwidth}
\caption{Damage effect comparison of different algorithms for various subnets. (\textbf{a})\,The subnet is ER random network. (\textbf{b})\,The subnet is BA scale-free network. (\textbf{c})\,The subnet is Goh scale-free network.\label{fig_damage}}
\end{figure*}

The model of Equation~(\ref{eq_HX}) can be solved by a simple linear programming algorithm. {\color{black}After obtaining the set of critical nodes based on the importance index,} combat networks of different models are damaged according to the map between encoding and decoding results. In the experiment, the damage intensity ranges from 1 to 20 and the damage simulation of each type of model network is repeated 100 times. The average results of the experiments are shown in Figure~\ref{fig_damage}. The shaded area in the figure represents the 90\% confidence interval.

\begin{figure*}[!htbp]
\centering
  \begin{minipage}{1\textwidth}
    \includegraphics[width=6cm]{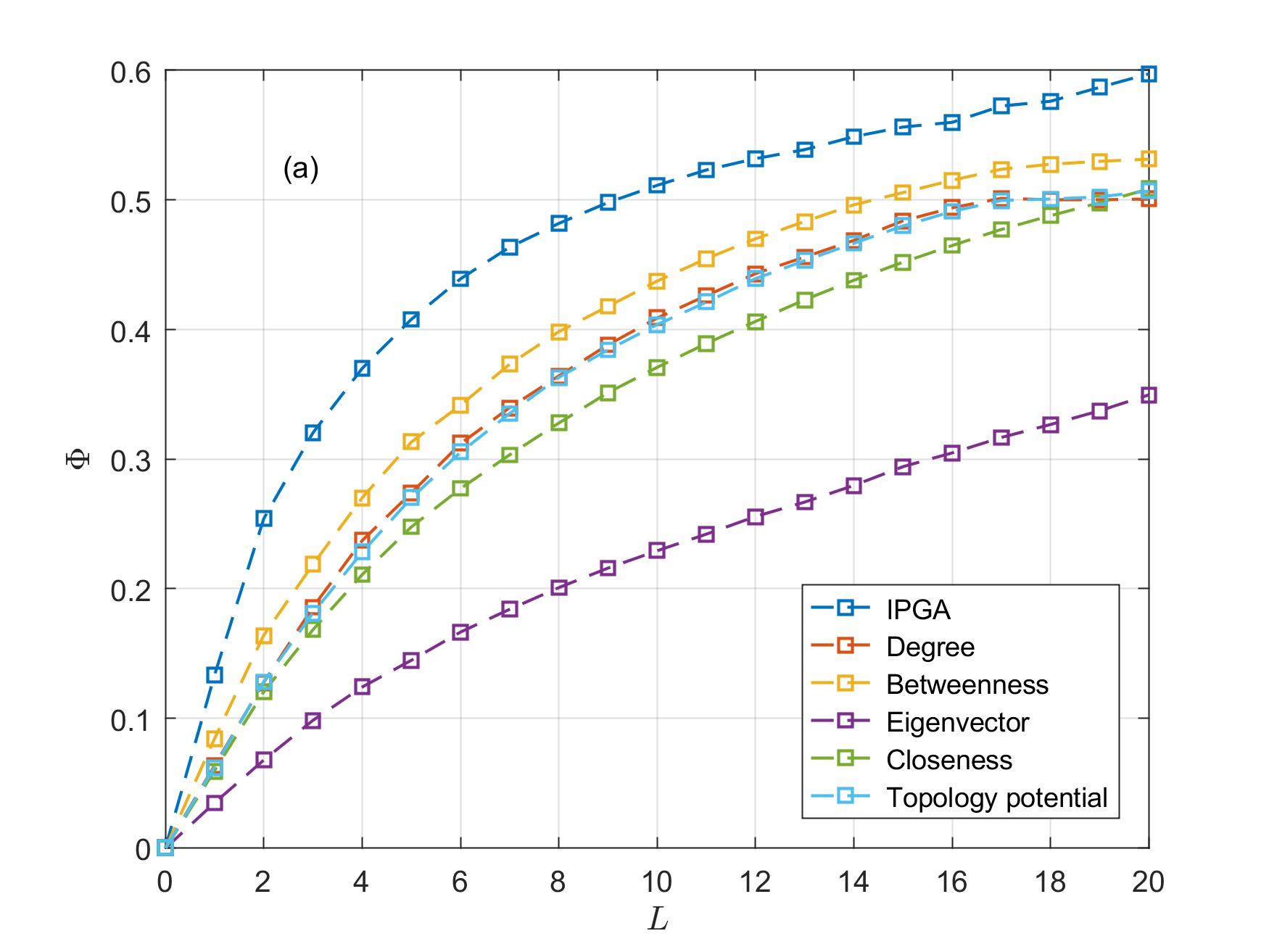}
    \includegraphics[width=6cm]{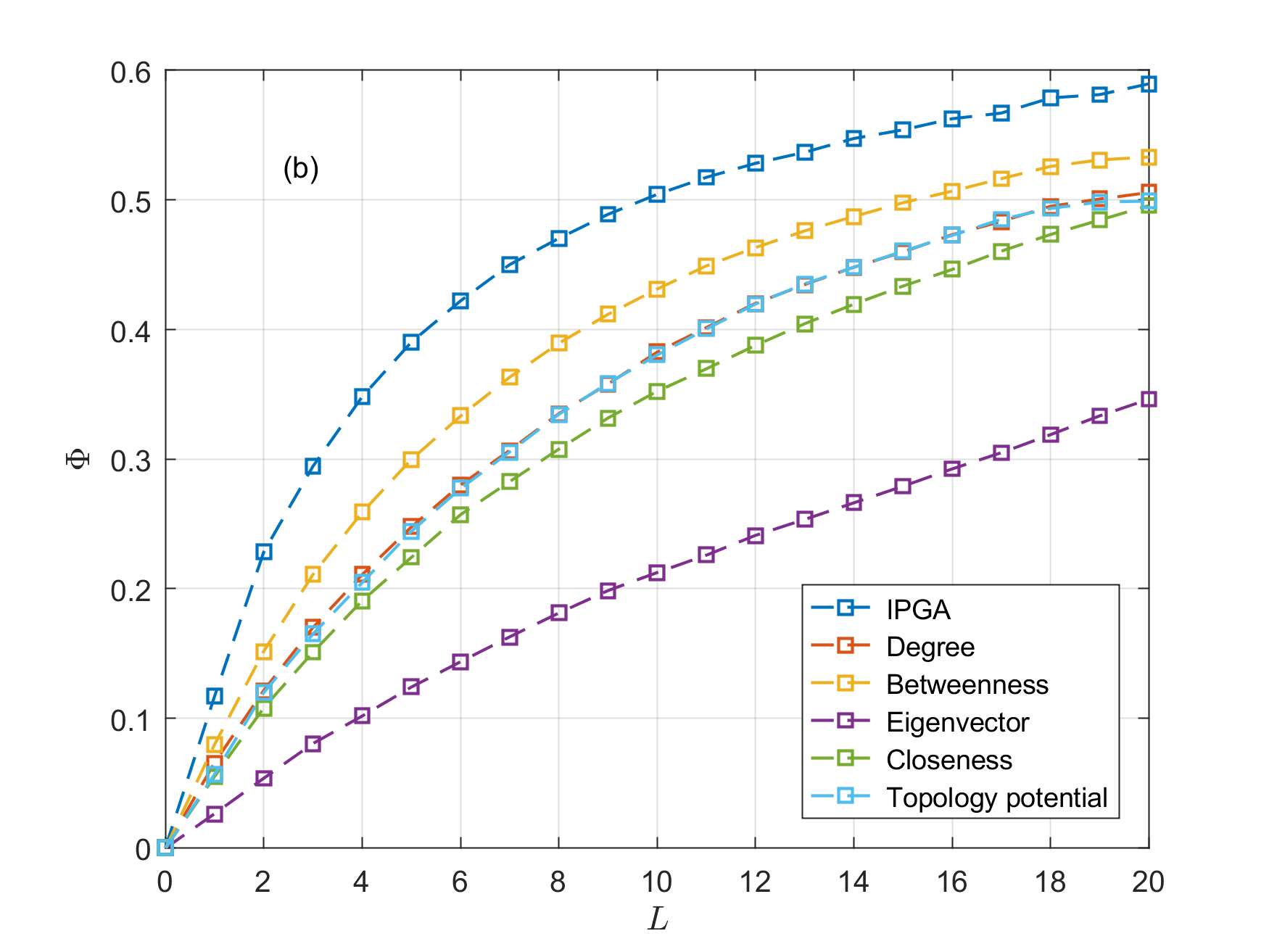}
    \includegraphics[width=6cm]{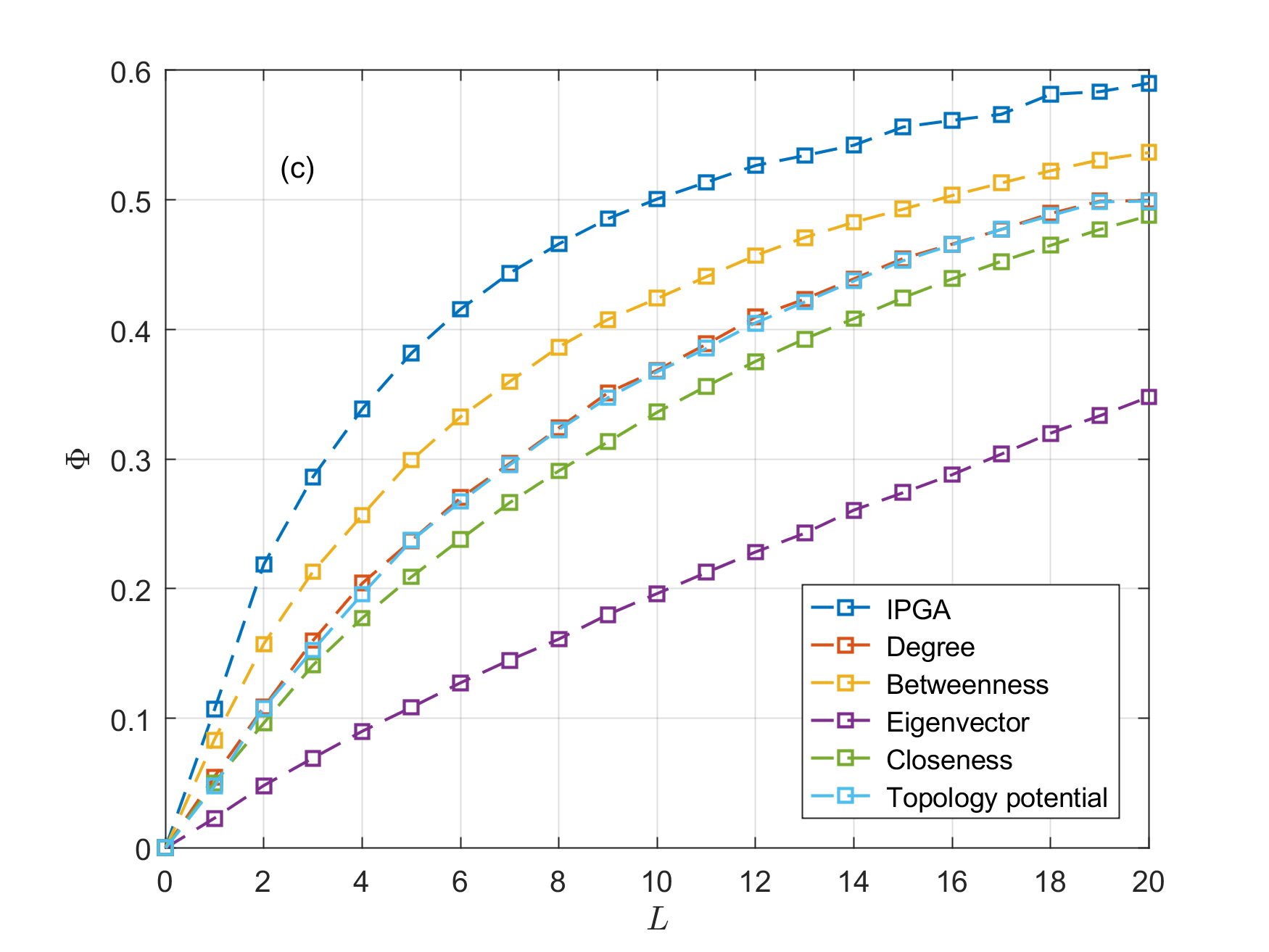}
  \end{minipage}

  \begin{minipage}{1\textwidth}
    \includegraphics[width=6cm]{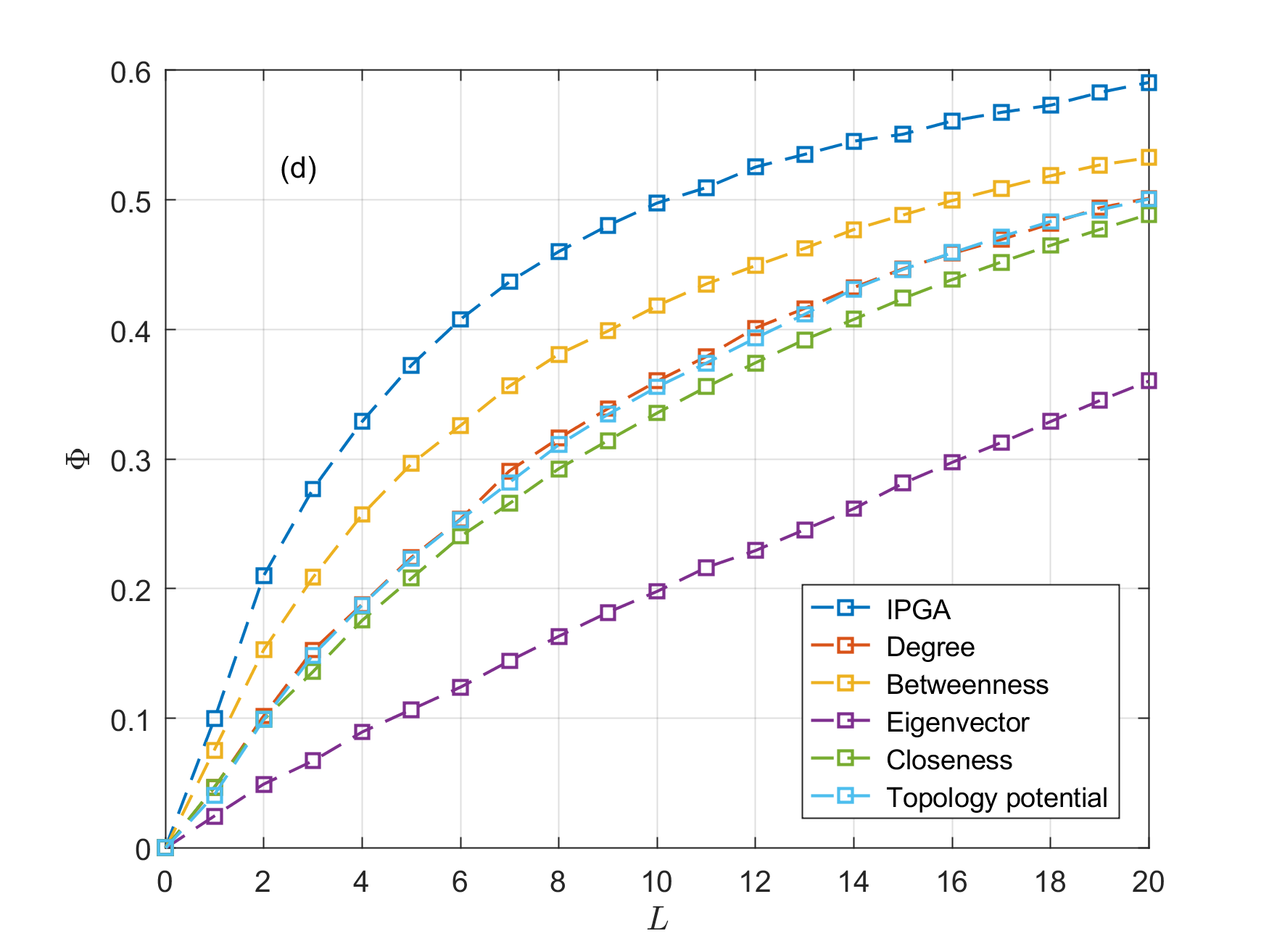}
    \includegraphics[width=6cm]{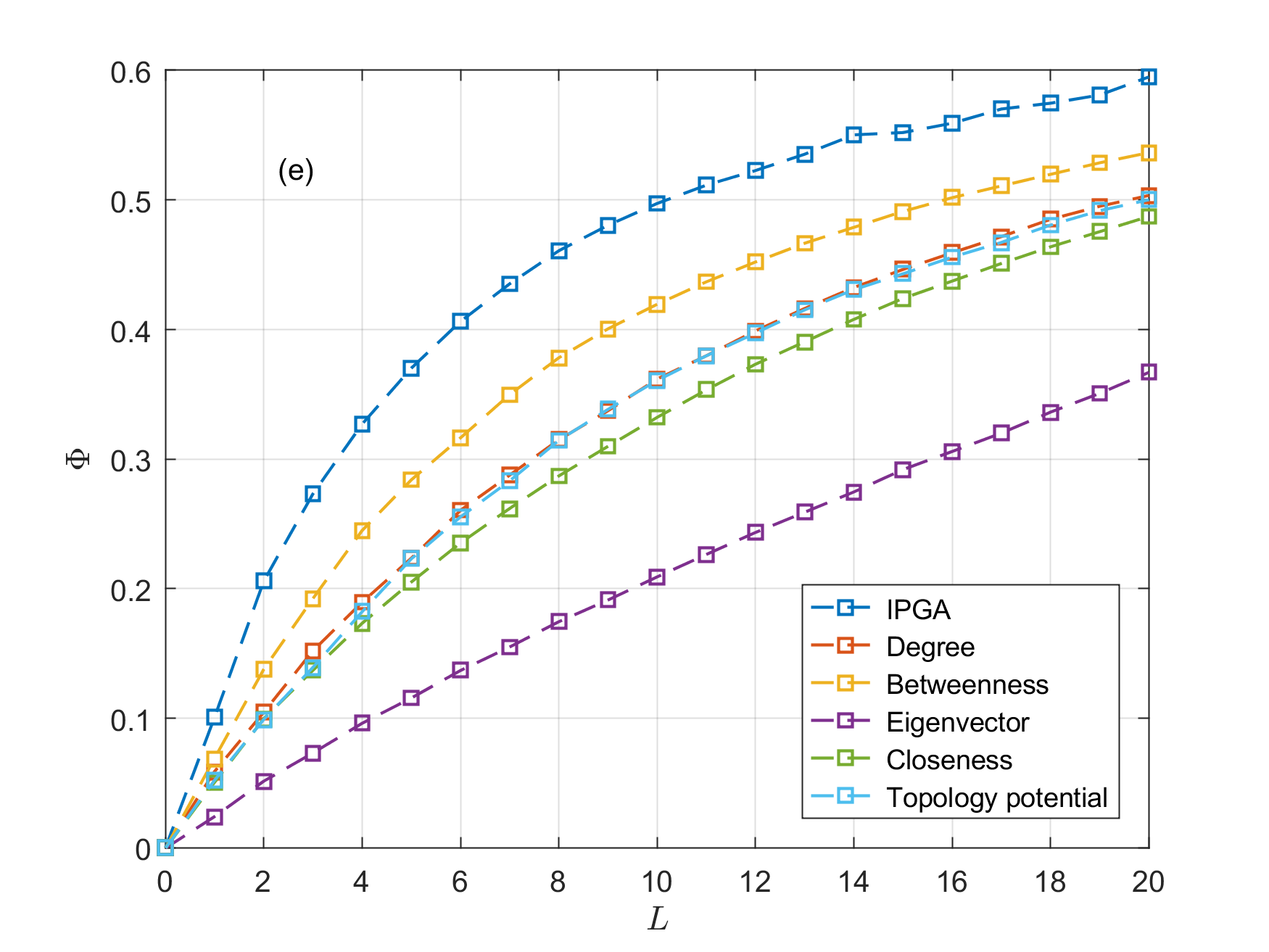}
    \includegraphics[width=6cm]{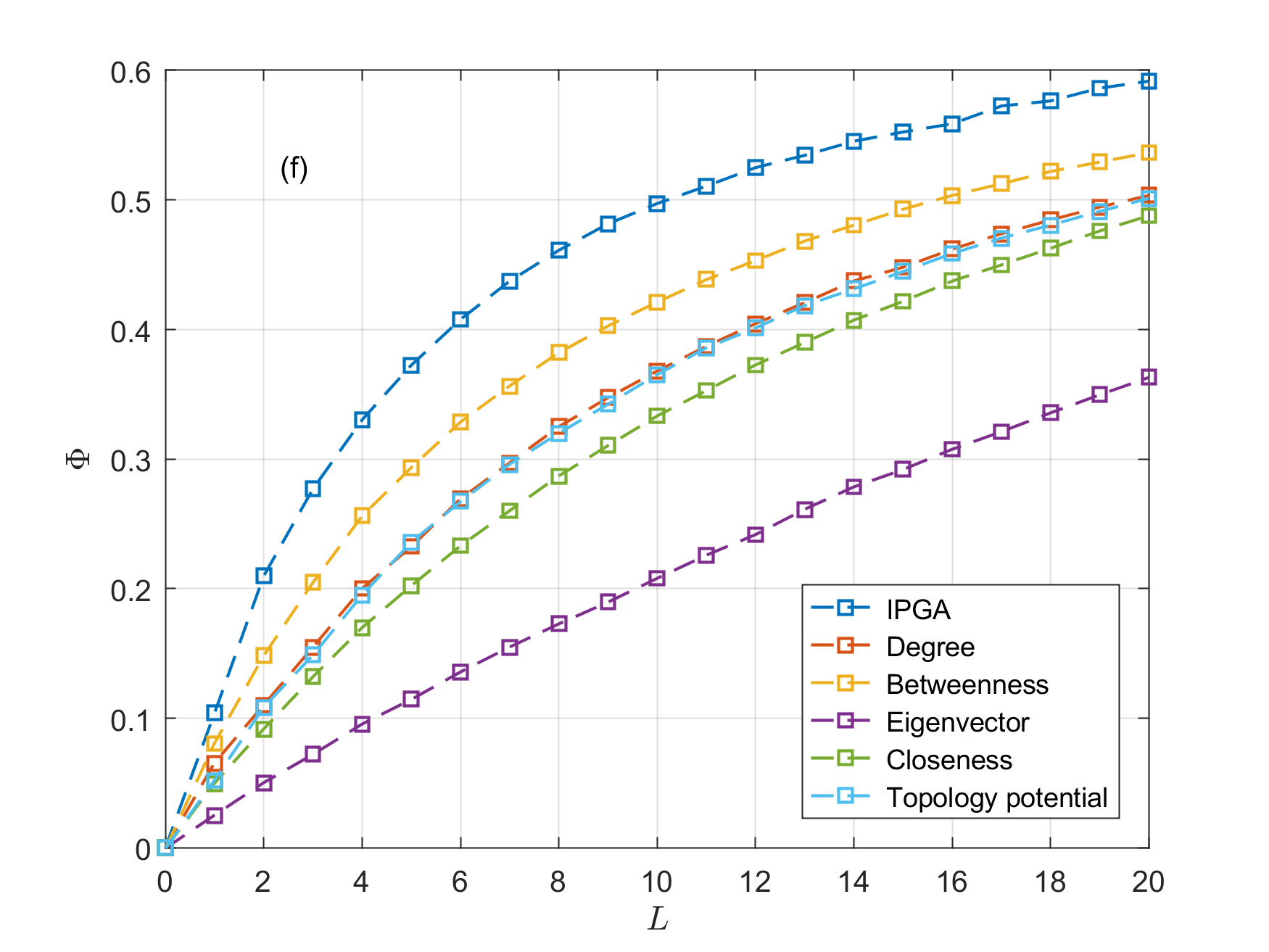}
  \end{minipage}
\caption{ Damage effect comparison of different algorithms for various
Goh subnets with adjustable parameter $\beta$. (\textbf{a}) $\beta = 2.5$. (\textbf{b}) $\beta = 3$. (\textbf{c}) $\beta = 3.5$. (\textbf{d}) $\beta = 4$. (\textbf{e}) $\beta = 4.5$. (\textbf{f}) $\beta = 5$. \label{fig_para_beta}}

\end{figure*}

%It can be seen from the figure that whether for the ER random network or the scale-free network model, the damage effect by attacking the nodes set obtained by the IPGA is the best, and has greater advantages over other algorithms.
\textcolor{black}{It can be seen from the figure that whether for the ER random network or the scale-free network model, the IPGA can find out the nodes set with the best damage effect on a combat network, and this method has great advantages over other algorithms.}
For the damage effect of algorithms based on degree, betweenness, and topological potential, there is not much difference in general. The {\color{black}result of} degree-based algorithm in ER random network is better than that based on betweenness and topological potential, and the results of the latter two are very similar. In the scale-free network, the algorithm based on betweenness is better than the algorithms based on degree and topological potential, and the damage effects of the latter two are also very similar. The {\color{black}result of} algorithm based on closeness is a little worse than {\color{black}that} based on degree, betweenness, and topological potential, but as the increase of damage intensity, the damage effects of those {\color{black}algorithms} are getting closer and closer. The algorithm based on eigenvector is the worst, and the measure of damage effect is always small. In summary, it is feasible and effective to identify the critical nodes set by damaging the combat network with limited costs based on the IPGA. According to the confidence interval, we can see that the above conclusion is clear and credible, so the confidence interval will not be annotated in the following experiments.

\subsubsection{Comparison with Different Network Parameter $\beta$}

\begin{figure}[!htbp]
\centering
  \includegraphics[width=8cm]{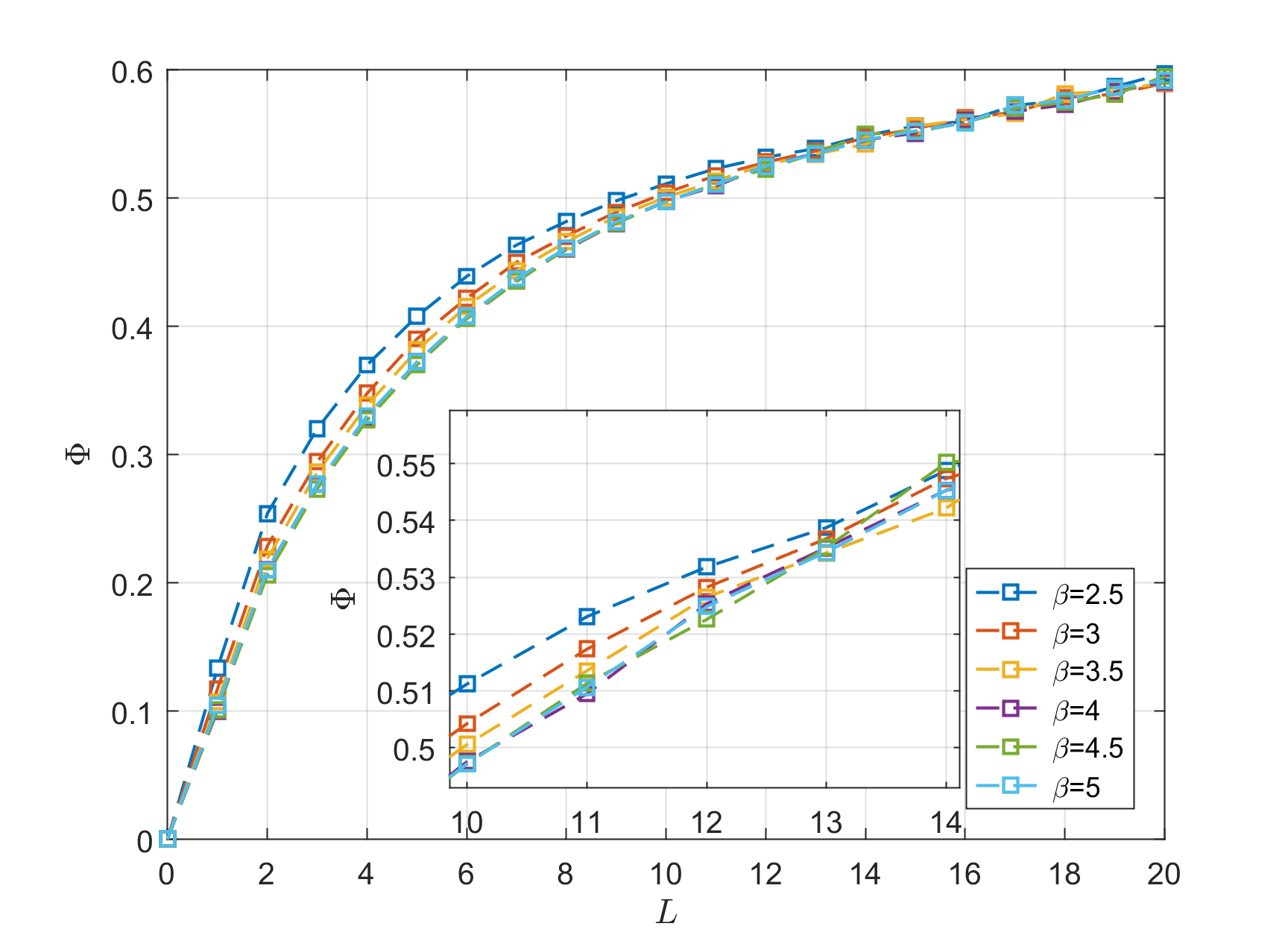}
\caption{Damage effect comparison for the IPGA with different parameter $\beta$ .\label{fig_allIPGA}}
\end{figure}

Simulation experiments based on the Goh subnet, whose power law parameter is configurable, are conducted to examine the effects of the various methods under different network characteristics. The subnet's parameter $\beta$ ranges from 2.5 to 5 {\color{black}with an interval of 0.5.} The result is shown in Figure~\ref{fig_para_beta}. From the figure, it can be seen that the IPGA continues to be the optimal approach for maximizing the damage effect of combat networks under cost constraints. The effect comparison of the IPGA method under the different parameter $\beta$ is shown in Figure~\ref{fig_allIPGA}. In general, the damage effect of the IPGA tends to decrease as the parameter $\beta$ increases, but the zoomed-in plot shows that the effect fluctuates slightly after $\beta>4$, which may be due to the large parameter making the network structure uniformly distributed.

\subsubsection{Optimal Gap Effect for Reduced Size Networks}
\begin{figure}[!htbp]
\centering
  \includegraphics[width=8cm]{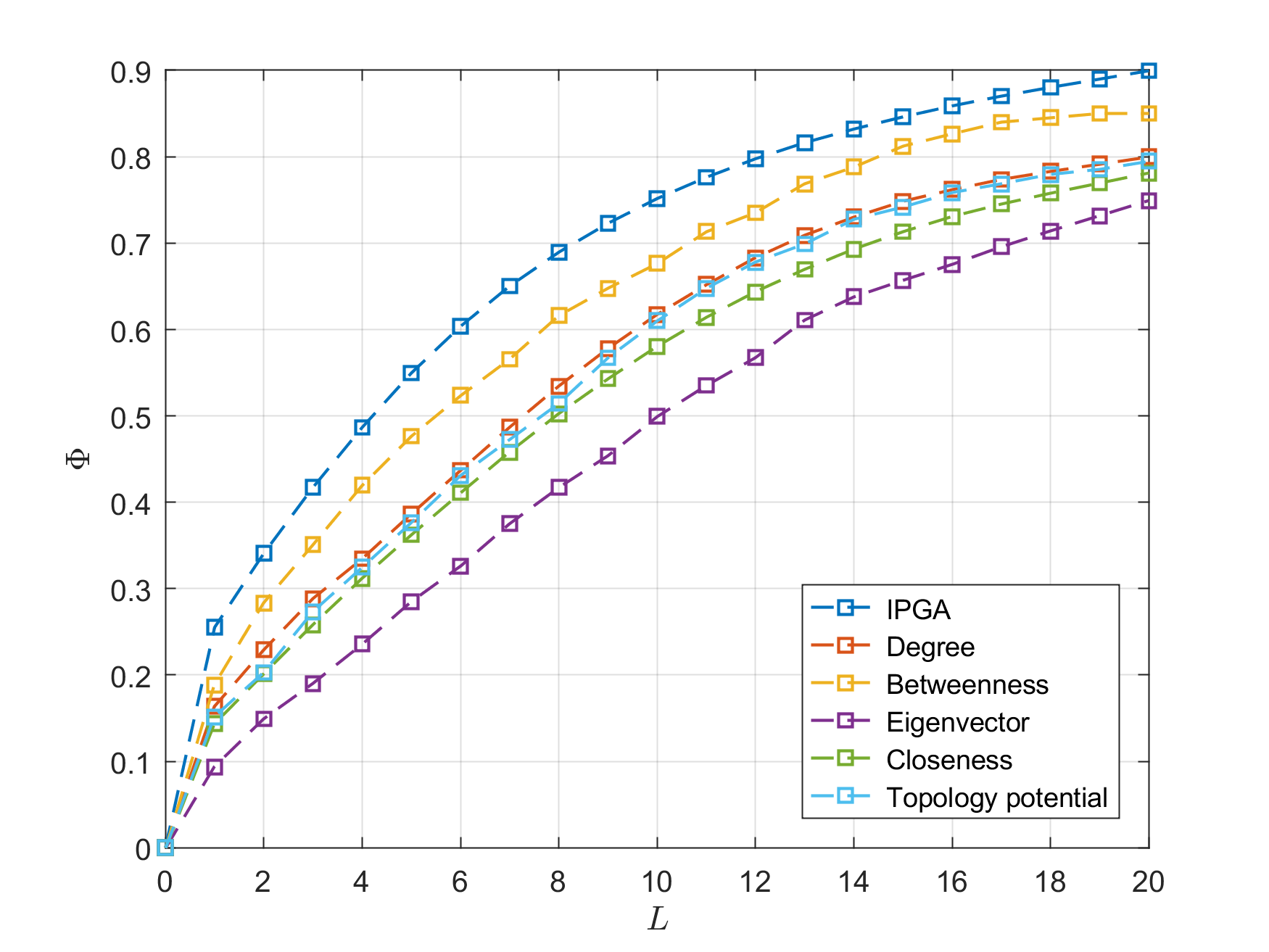}
\caption{Damage effect comparison on the Goh subnet with a scale of 70 nodes.\label{fig_70node}}
\end{figure}

\begin{figure}[!htbp]
\centering
  \includegraphics[width=8cm]{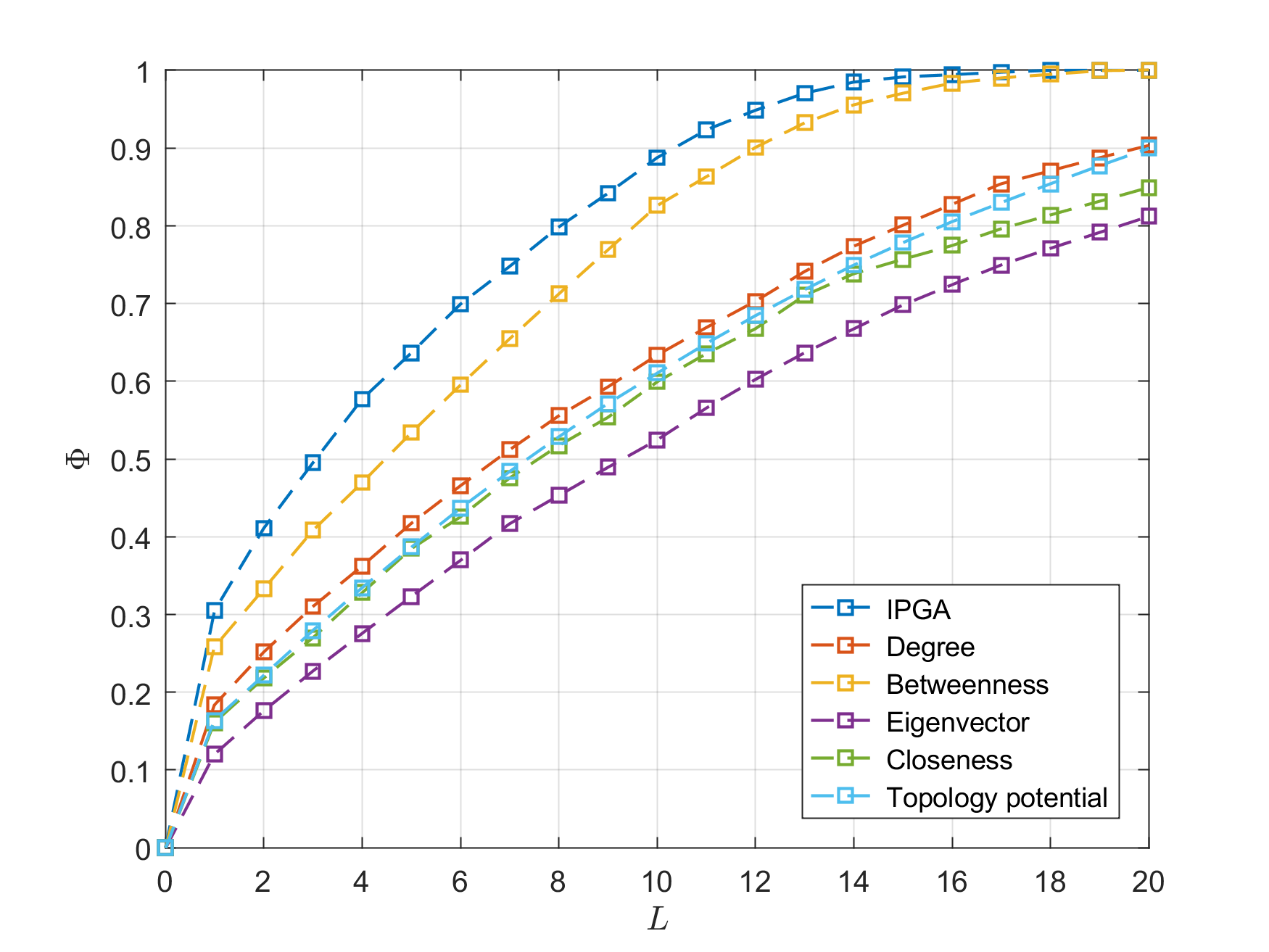}
\caption{Damage effect comparison on the Goh subnet with a scale of 50 nodes.\label{fig_50node}}
\end{figure}

When the size of the combat network is reduced, the same parameters {\color{black}are used as in the first experiment}, and only the number of nodes is changed to generate the network for simulation experiments. Here we still adopt the Goh subnet as the model network.  Figure~\ref{fig_70node} and Figure~\ref{fig_50node} give a comparison of the damage effect of the different methods when the number of network nodes is 70 and 50, respectively. From the figure, {\color{black}we can see that as the size of the combat network decreases,} the gap among optimal solutions of all methods is getting closer. In addition, the optimality gap of the IPGA {\color{black}appears when the node size is 50,} which indicates that the optimal solution of the IPGA and method based on betweenness is the same. If the network size is further reduced, all methods will achieve the same optimal result. However, the actual combat network has a large number of nodes and edges, so the problem of this optimality gap will not arise.


\subsubsection{Attack Law of Combat Network Damage Maximization}
\begin{figure}[!htbp]
\centering
  \includegraphics[width=8cm]{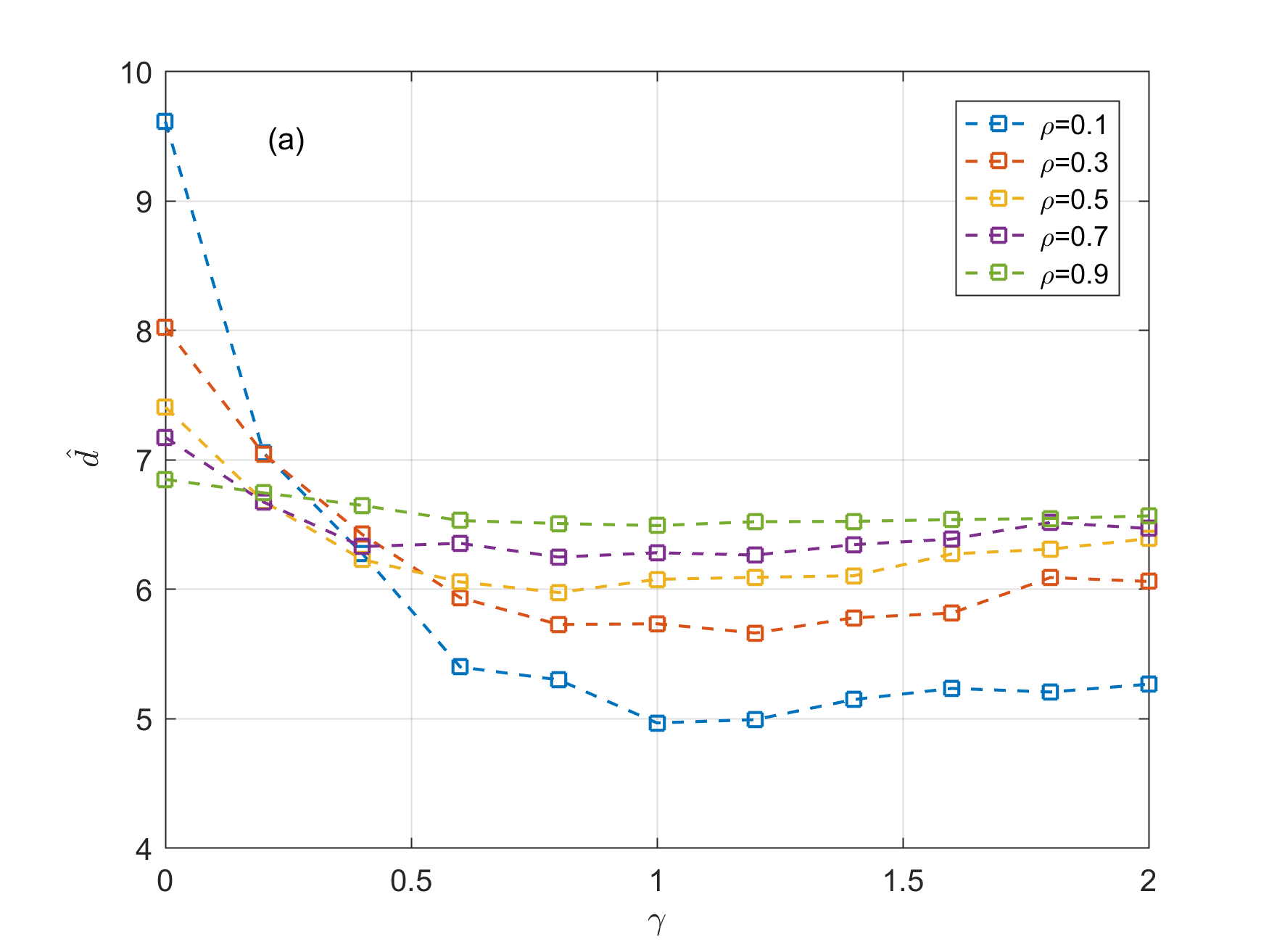}\\
  \includegraphics[width=8cm]{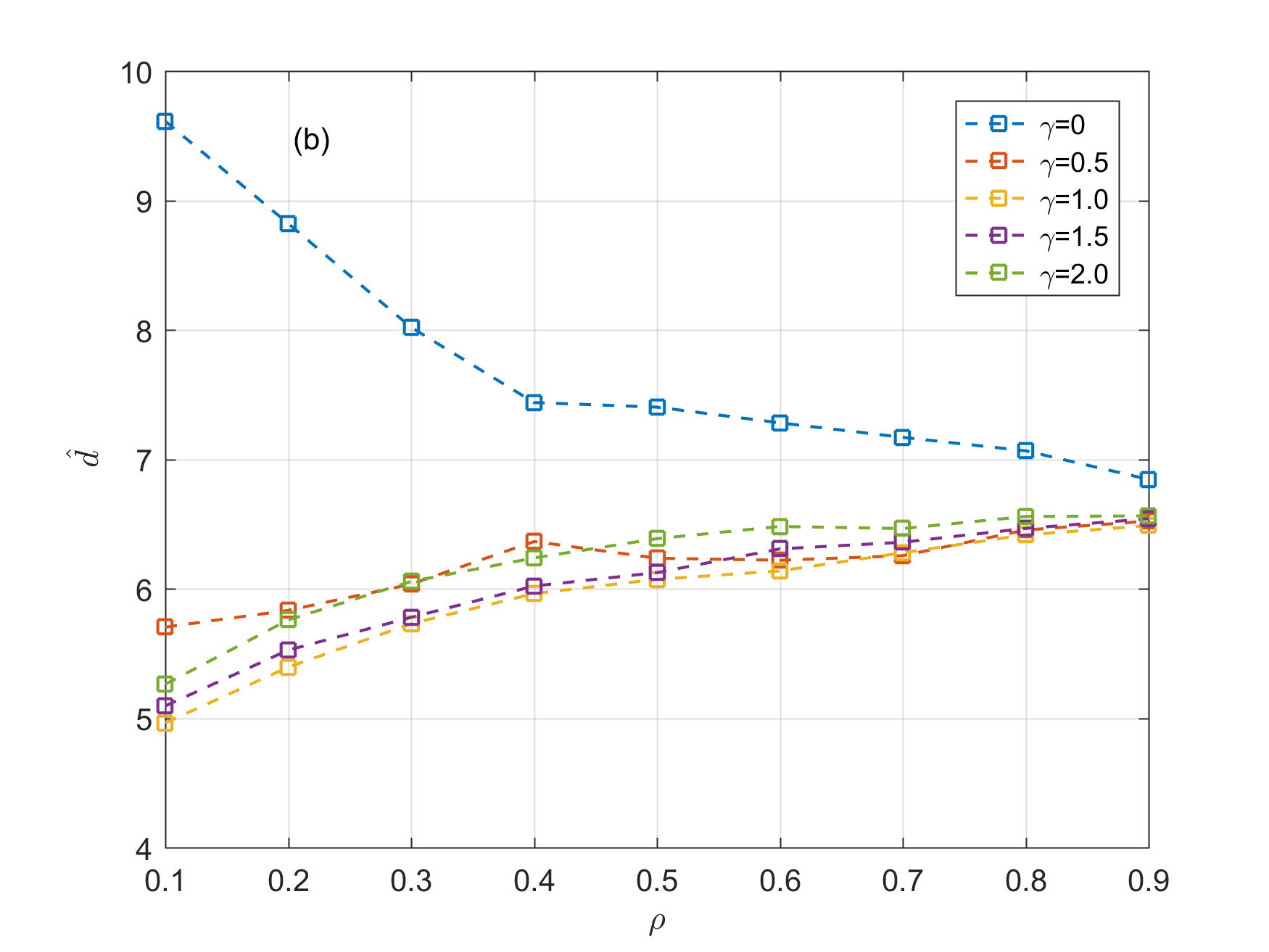}\\
  \includegraphics[width=8cm]{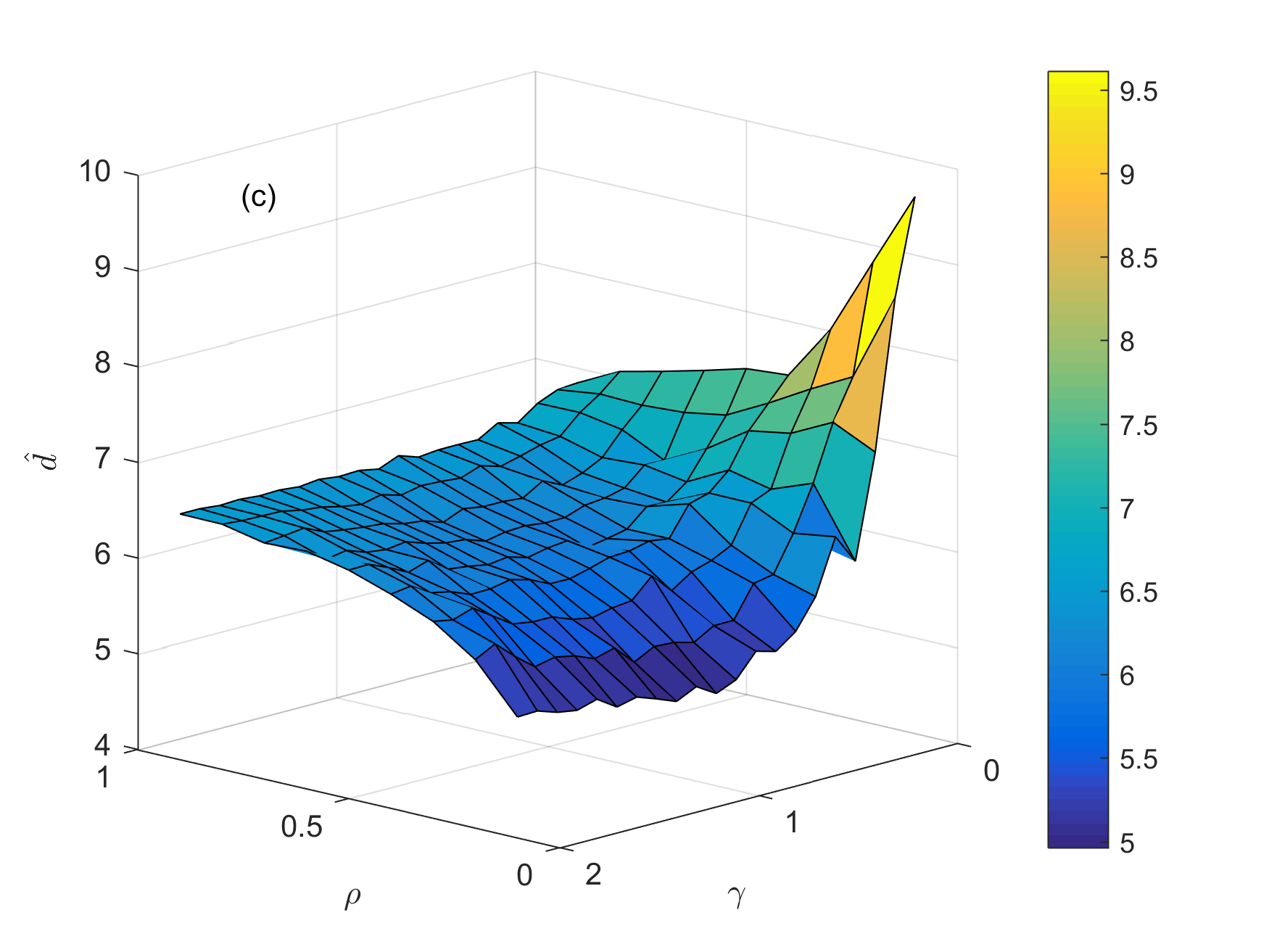}
  \caption{\color{black}Relations among the average degree $\hat{d}$, cost constraint ratio $\rho$, and cost power parameter $\gamma$. (\textbf{a}) The relation between $\hat{d}$ and $\gamma$. (\textbf{b}) The relation between $\hat{d}$ and $\rho$. (\textbf{c}) The 3D view of the relation among $\hat{d}$, $\rho$, and $\gamma$.\label{fig_dyp}}
\end{figure}

In order to explore the attack law of damage maximization problem of combat networks based on the IPGA, the Goh scale-free network is used as the model network for simulation. The damage intensity in the experiment is the damage intensity when the upper limit of the cost {\color{black}constraints is reached. And the simulation results are the mean results} based on 100 Goh networks. When the cost power parameter $\gamma$ varies from 0 to 2, the average degree $\hat{d}$ of the damaged nodes set is shown in Figure~\ref{fig_dyp}(a). It can be seen from the figure that with the {\color{black}increase} of $\gamma$, the curves of the average degree corresponding to different cost constraints ratio parameters $\rho$ basically show a trend of first decreasing and then stabilizing. {\color{black}When $\gamma$ is small, the node with a large degree is more likely to be attacked.} As $\gamma$ increases, the average degree of attacked nodes set gradually decreases and keeps stable. In addition, the looser the cost constraints, the higher the average degree of the attacked nodes set.

When the constraint ratio parameter of cost changes from 0.1 to 0.9, the value of the average degree of damaged nodes set is shown in Figure~\ref{fig_dyp}(b). It can be seen from the figure that except for the case where the cost power parameter is zero, the curves of average degree in other cases basically show a trend of first increasing and then approaching a certain value. When $\gamma =0$, the damage costs of all nodes are the same. {\color{black}So it is more inclined to attack the node with a large degree under the cost constraints. While in other cases, attacks on the node with a higher degree will not be performed,} unless the cost constraints are gradually released and resources are sufficient to allocate.

In order to better observe the above rules, the relation among the average degree of damaged nodes set, the cost power parameter, and the cost constraints ratio parameter is plotted in Figure~\ref{fig_dyp}(c). It is clear that the observed results are consistent with the previous description.

\begin{figure}[!htbp]
\centering
  \includegraphics[width=8.5cm]{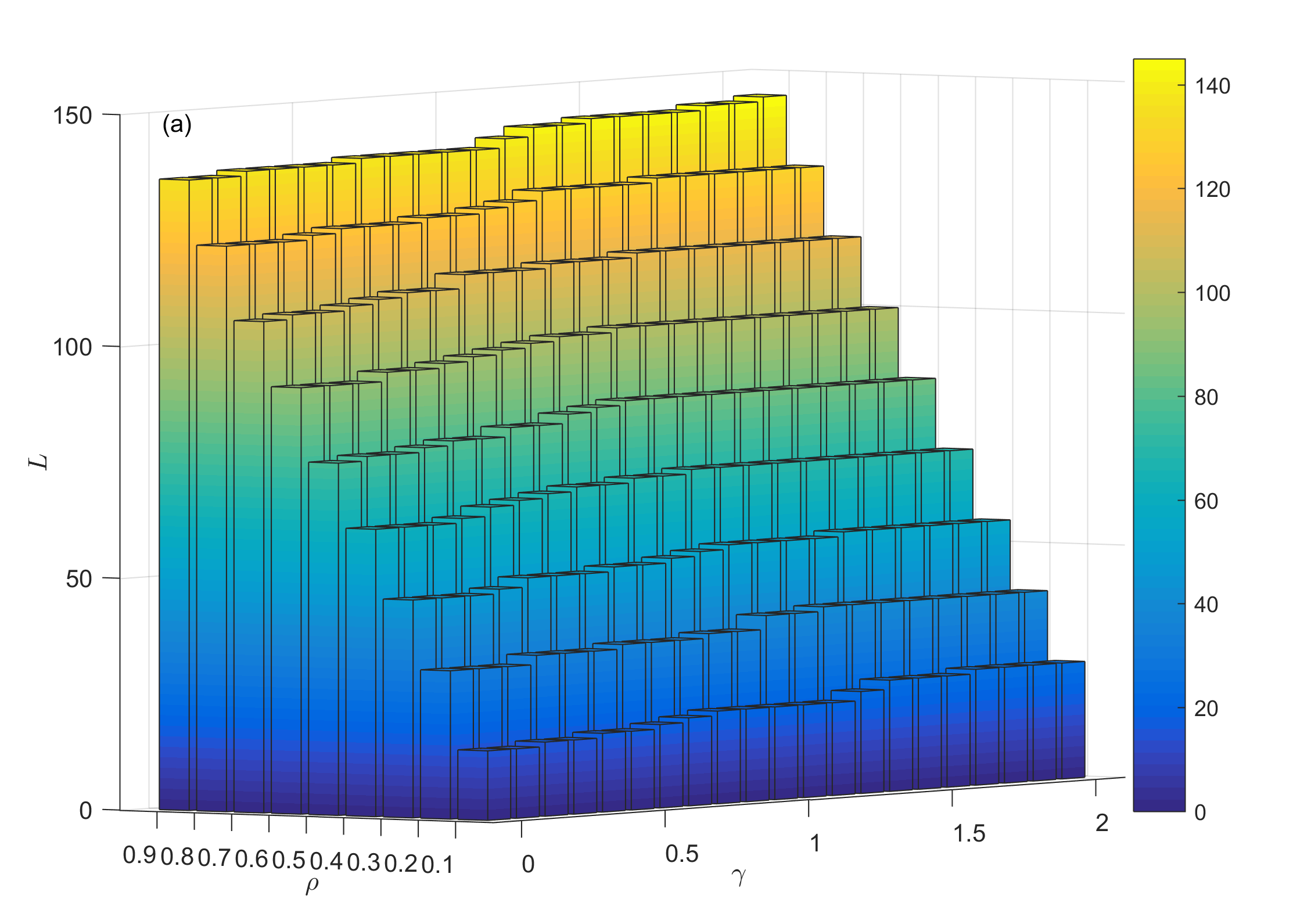}\\
  \includegraphics[width=8cm]{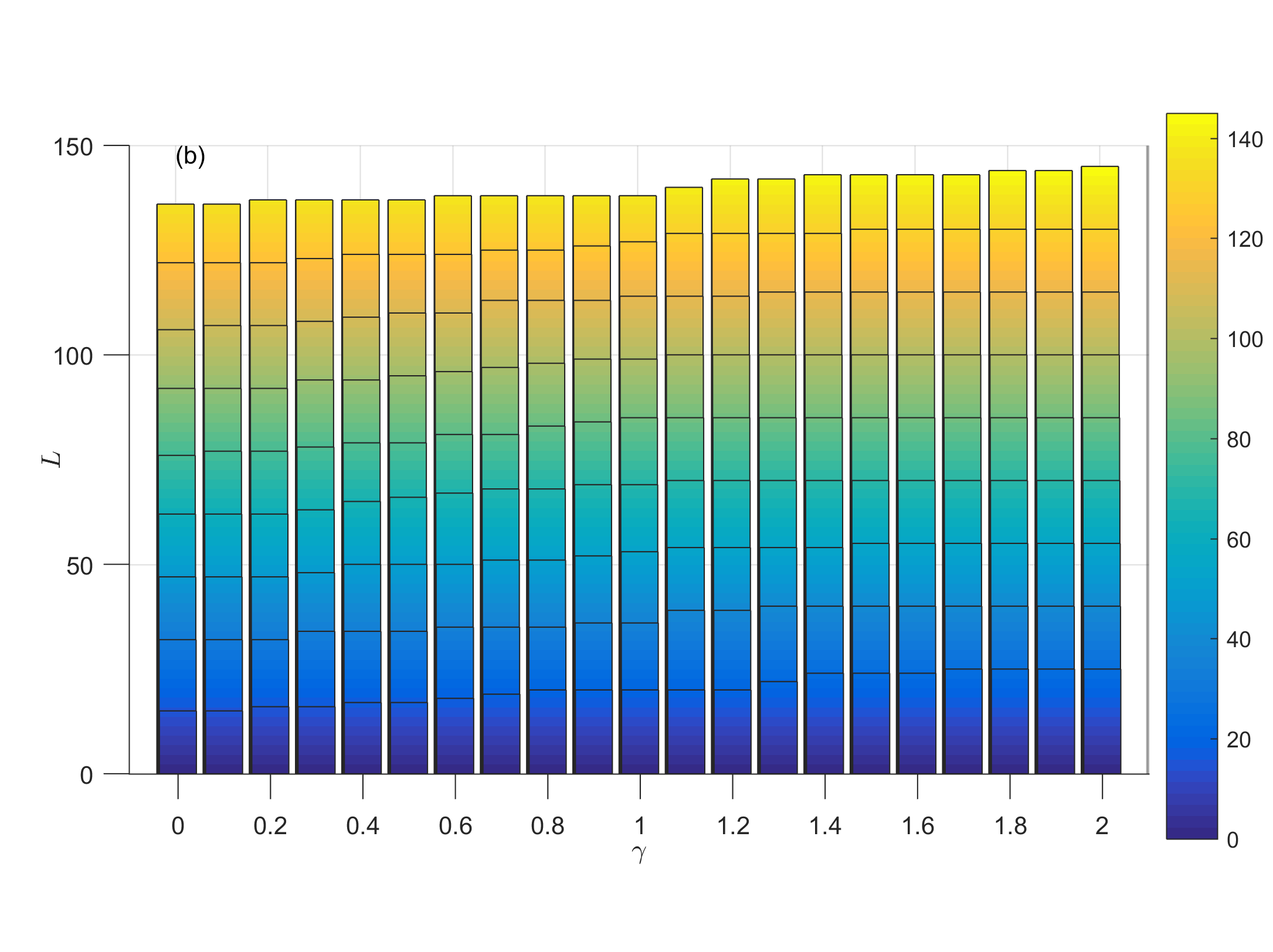}
\caption{Relations among damage intensity $L$ and parameters $\gamma$ and $\rho$. (\textbf{a}) Damage intensity $L$ with different $\gamma$ and $\rho$. (\textbf{b}) Front view of sub-figure~(a). \label{fig_L}}
\end{figure}

Apart from the average damage degree being affected by $\gamma$ and $\rho$, the damage intensity $L$ {\color{black}for the combat network} is also inseparable from them. The relations among $L$, $\gamma$, and $\rho$ are shown in Figure~\ref{fig_L}. It can be deduced that with the increase of the parameter $\rho$, the damage intensity increases significantly, while the increase of $\gamma$ has little effect on the damage intensity, {\color{black}indicating that the damage effect of the combat network with limited costs mainly depends} on the upper bound of the damage cost, and has little correlation with the damage cost of a single node. Table~\ref{tab_var} presents the variance of $L$ as the increasing of $\rho$. It can be known from the table that the variance of damage intensity is small, and with the increase of $\rho$, the fluctuation of damage intensity becomes smaller and smaller, which further verifies the above point of view.

\begin{table}[!htbp]
\centering
\footnotesize{
\caption{Variance of $L$ with the increase of $\rho$.\label{tab_var}}
	\begin{tabular}{cccccc}
	\toprule
	$\rho$ & 0.1 & 0.2 & 0.3 &0.4 & 0.5\\
	\hline
	Var($L$) &1.5224 & 0.8269 & 0.3921 & 0.1846 & 0.1223 \\
	\hline
	$\rho$ & 0.6 & 0.7 & 0.8 & 0.9&\\
	\hline
	Var($L$) & 0.0791 & 0.0555 & 0.0305 & 0.0101&\\
	\bottomrule
	\end{tabular}}
\end{table}

\subsection{Algorithmic Convergence And Computation Complexity Analysis}
\begin{figure}[htbp]
\centering
\includegraphics[width=8cm]{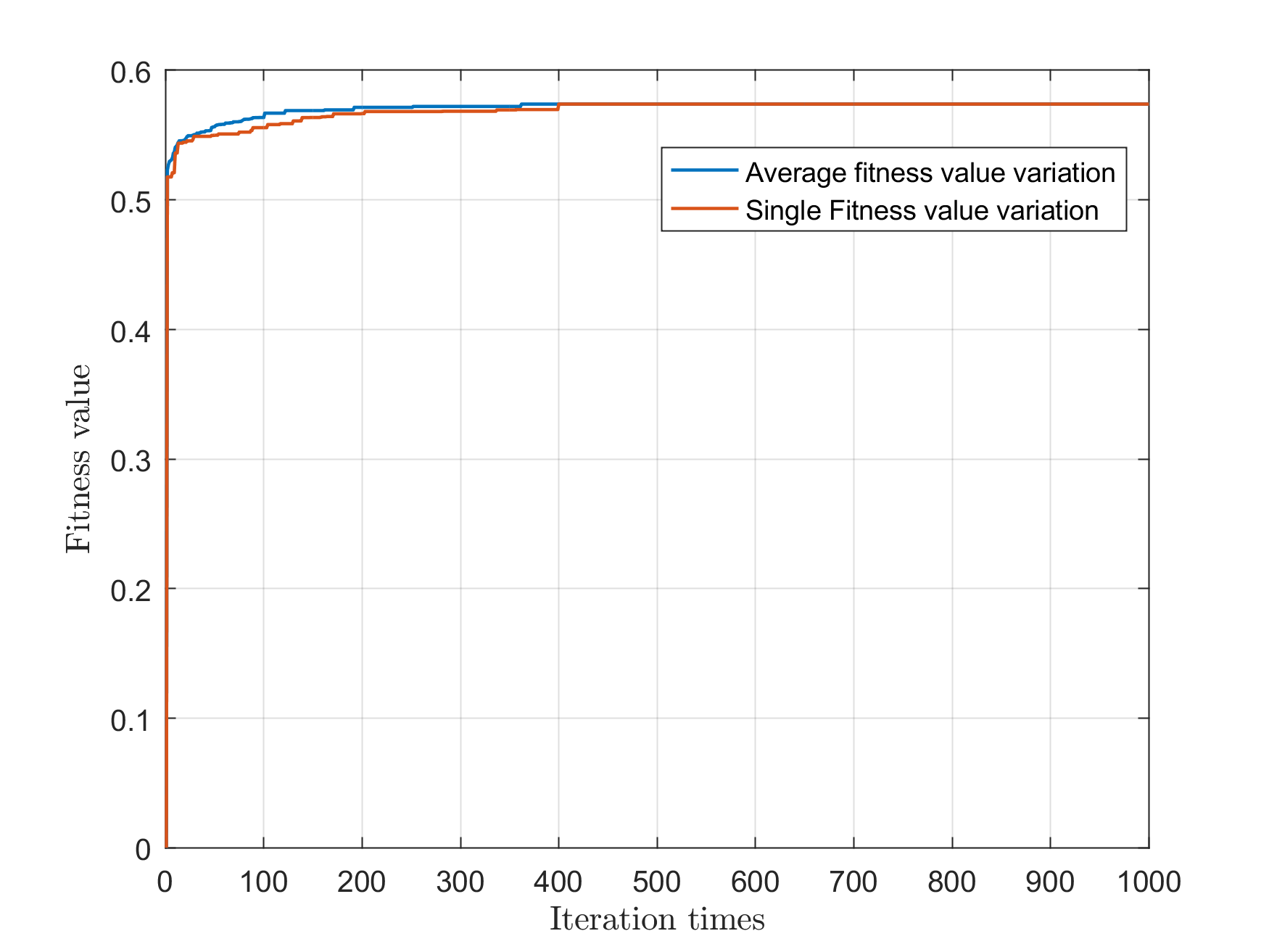}
\caption{Convergence curve of the IPGA. \label{fig_convergence}}
\end{figure}

{\color{black}For the optimal problem with GA algorithm}, the convergence of the algorithm is supposed to analyze to strengthen the confidence of the results. Chang et al.~\cite{b31} proved the convergence of the GA algorithm for network topology both locally and globally. On the basis of this, we verify the convergence of the proposed algorithm through multiple simulations. Figure~\ref{fig_convergence} {\color{black}shows the convergence curves for damaging 18 nodes in a combat network based on the Goh subnet with a total scale of 150 nodes,} where the average convergence curve is shown in blue and the convergence curve for a particular experiment is shown in red. As can be seen from the figure, the algorithm converges after about 400 iterations for both the result of a single experiment and the average result of multiple experiments, indicating that the IPGA algorithm can converge to the optimum, and the theoretical and experimental results are consistent.

\begin{figure}[!htbp]
\centering
\includegraphics[width=9cm]{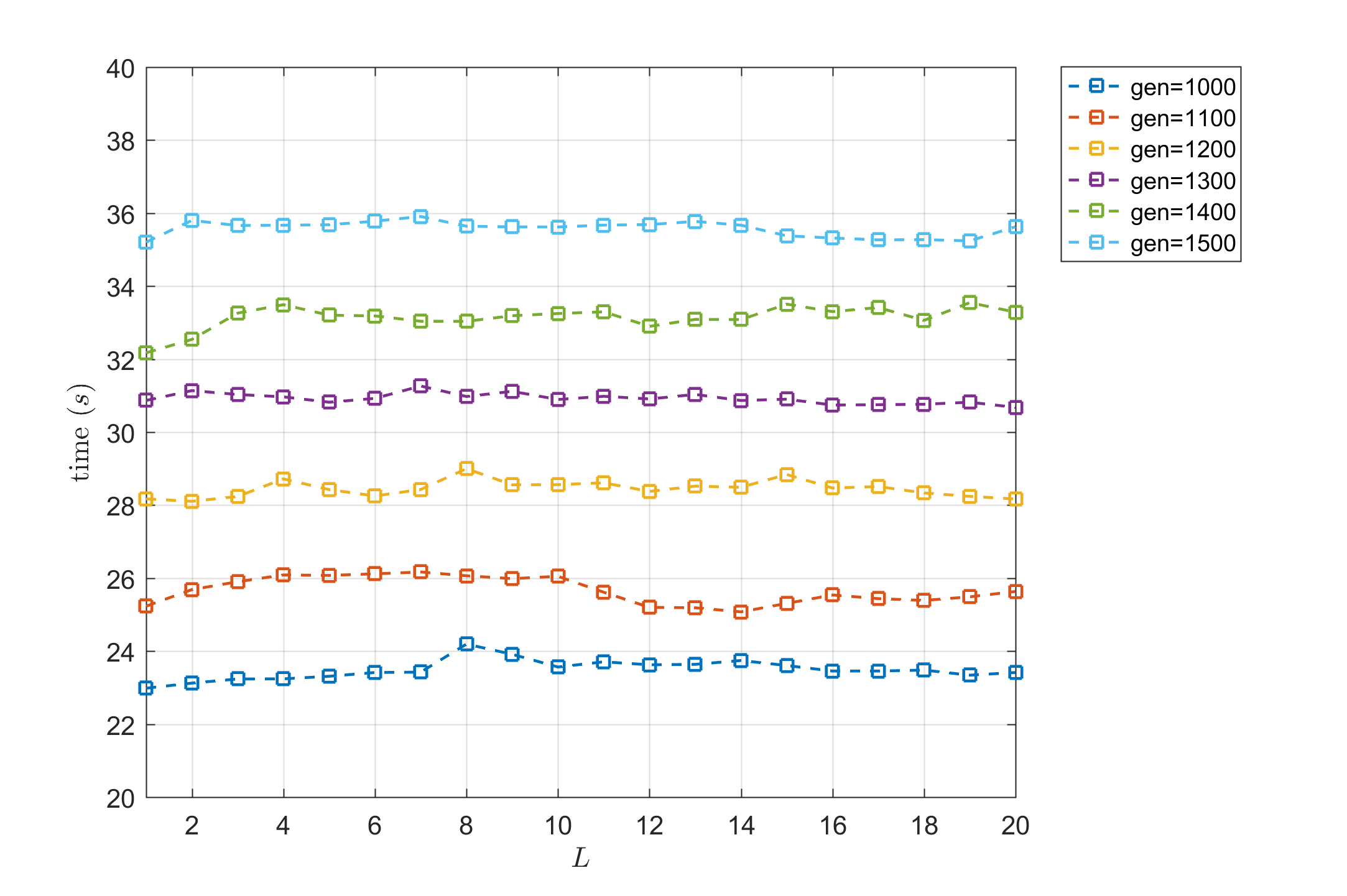}
\caption{Calculation time simulation of the IPGA. \label{fig_time}}
\end{figure}

As for the computation complexity of the IPGA, by analyzing the calculation process and related parameters, it can be obtained that the complexity of fitness assessment {\color{black}mainly depends} on the calculation of the maximal component and the number of attack links. The complexity of the former according to the Breadth First Search (BFS) or Depth First Search (DFS) is $O(N(N+M))$, and the complexity of the latter is $O({N_\rm{O}^4}{ N_\rm{P}^2}{N_\rm{D}^2}{N_\rm{A}^2})$ at most. The damage strength is $L$, but it is only involved in the encoding and decoding constraints and not in other calculations. During the crossover and mutation process, the complexity of them are $O(\frac{N_{\rm p}}{2} NP_{\rm c})$ and $O(N_{\rm p} NP_{\rm m})$, respectively. On the basis of the above analysis and the procedure of the IPGA, combined with the {\color{black}population} size $N_{\rm p}$ and iteration times $gen$,
it can be deduced that its computational complexity is $O(N_{\rm p} gen ( N(N+M)+{N_\rm{O}^4}{ N_\rm{P}^2}{N_\rm{D}^2}{N_\rm{A}^2}+\frac{N_{\rm p}}{2} NP_{\rm c}+N_{\rm p} NP_{\rm m}))$. That is, the complexity of the IPGA is $O(N_{\rm p} gen ( N(N+M+N_{\rm p})+{N_\rm{O}^4}{ N_\rm{P}^2}{N_\rm{D}^2}{N_\rm{A}^2}))$ at the most{\color{black}.
So} the complexity of the algorithm is only related to the size of the network, the population size and iteration times of the IPGA, but has nothing to do with the damage intensity. Theoretically, when the size of the network, the population size, the iteration times, and other parameters except the damage intensity remain constant, the running time of the IPGA will largely remain stable. Figure~\ref{fig_time} shows the relation between the calculation time of the IPGA and the damage intensity with the increase of iteration times. It can be seen from the figure that the calculation time does not vary greatly with the increase of damage intensity. And when the number of iterations increases, the calculation time changes significantly, which shows that our theoretical analysis is correct.

\section{Conclusions}
\label{sec:5}
The research on maximizing damage in a combat network has clear practical significance and is very important for protecting or destroying important units in the combat SOS. 
%20221223

In this paper, by studying the damage maximization problem for combat networks with limited costs, two main contributions of this work can be summarized. On the one hand, a more realistic damage maximization model is established, and an effective optimization algorithm is proposed to solve the model. On the other hand, a reference for the formulation of the attack strategy in battle is provided by exploring the attack law in the combat network. According to this, combat units can be protected or destroyed in a targeted manner.
%20221223

{\color{black}However, there are still some limitations to our proposal. Our approach only consider the damage cost constraint for a fixed number of nodes, so it is a single objective optimization problem. In fact, the number of attacked nodes may not be so easy to determine under conditions of incomplete information. The multi-objective problem needs further study if we want both low damage costs and good network damage results, and the determination of the number of damage nodes is also a difficult issue worth exploring. Apart from this, there are two more points worth investigating in the future research.} One is to study the application in the actual combat network. The other is to improve the computing efficiency by optimizing the algorithm, so as to be applied to the larger-scale combat network damage problem.

\section{Appendix}
\begin{table}[H]
\caption{Acronym and Meaning. \label{tab_abbr}}
\centering
\begin{tabular}{p{2.5cm}p{5cm}}
  \toprule
  Acronym & Meaning	\\
  \midrule
  SOS     & system-of-system  \\ \hline
  IELK	& Intelligence effectiveness link  \\ \hline
  IELP   	& Intelligence effectiveness loop  \\ \hline
  IPGA	& Improved genetic algorithm  \\ \hline
  BFS		& Breadth First Search  \\ \hline
  DFS	    & Depth First Search  \\
  \bottomrule
\end{tabular}
\end{table}

 \begin{IEEEbiography}[{\includegraphics[width=1in,height=1.25in,clip,keepaspectratio]{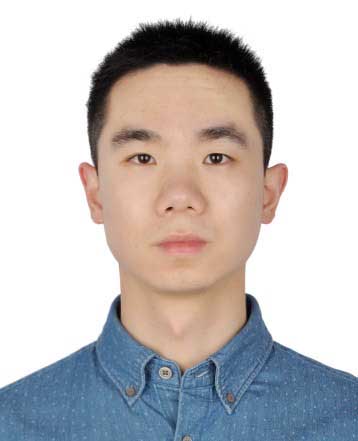}}]{Jintao Yu} was born in Zhuji, Zhejiang, China, in 1994. He received the M.S. degree in applied mathematics from Army Academy of Artillery and Air Defense, Hefei, China, in 2019.

 He is currently working toward the Ph.D. degree with operations research and systems engineering in Air Force Early Warning Academy, Wuhan, China. His research interests include complex network theory, intelligent optimization algorithms and their applications in practice.
 \end{IEEEbiography}

 \begin{IEEEbiography}[{\includegraphics[width=1in,height=1.25in,clip,keepaspectratio]{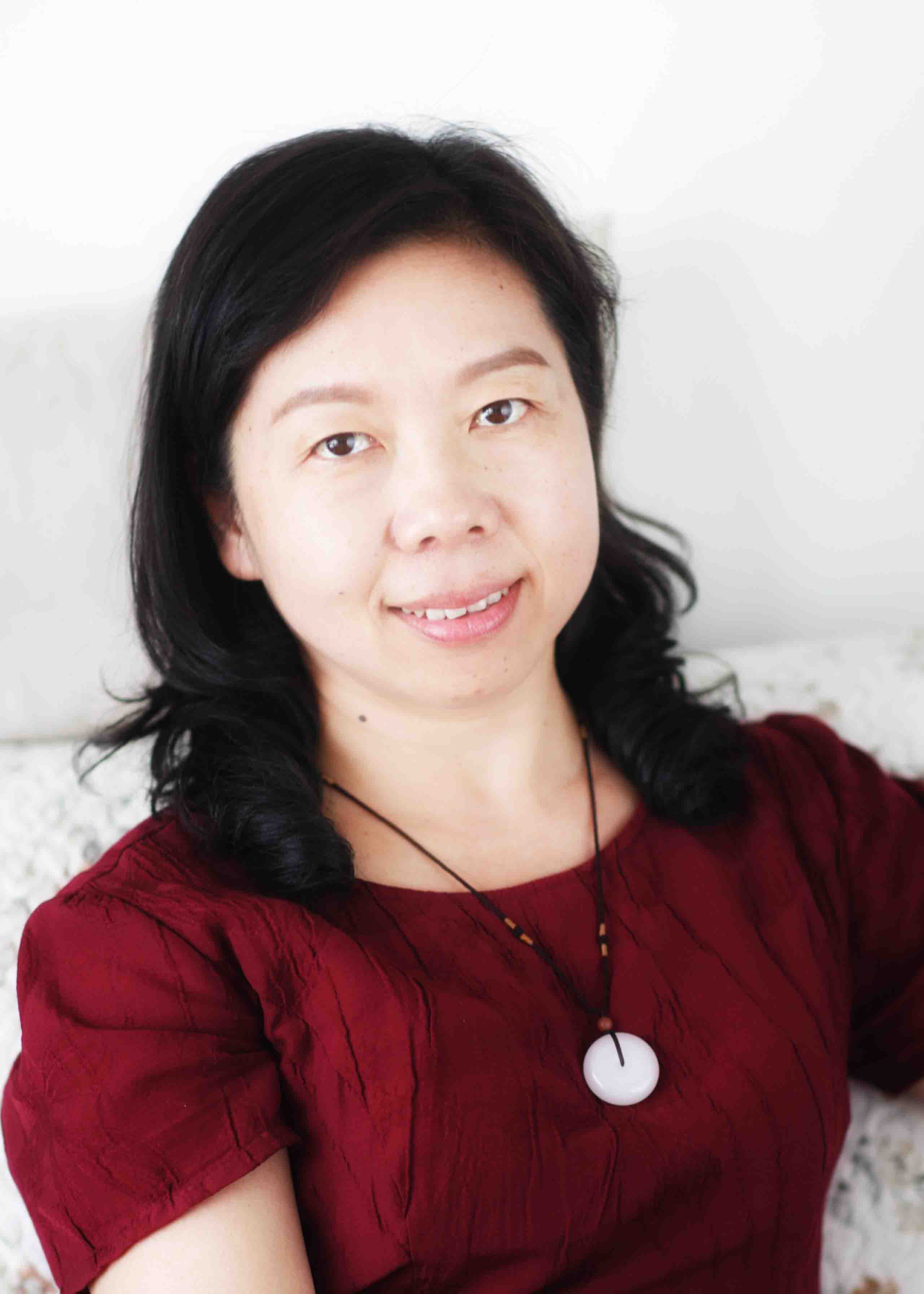}}]{Bing Xiao} was born in Bijie, Guizhou, China, in 1966. She received the Ph.D. degree in
 control theory and control engineering from Huazhong University of Science and Technology, Wuhan, China, in 2002.

 She is currently the Professor of the fourth department with Air Force Early Warning Academy and the board member of China system engineering society. Her research interests include system integration and analysis, complex network and multi-agent simulation modeling.
 \end{IEEEbiography}

 \begin{IEEEbiography}[{\includegraphics[width=1in,height=1.25in,clip,keepaspectratio]{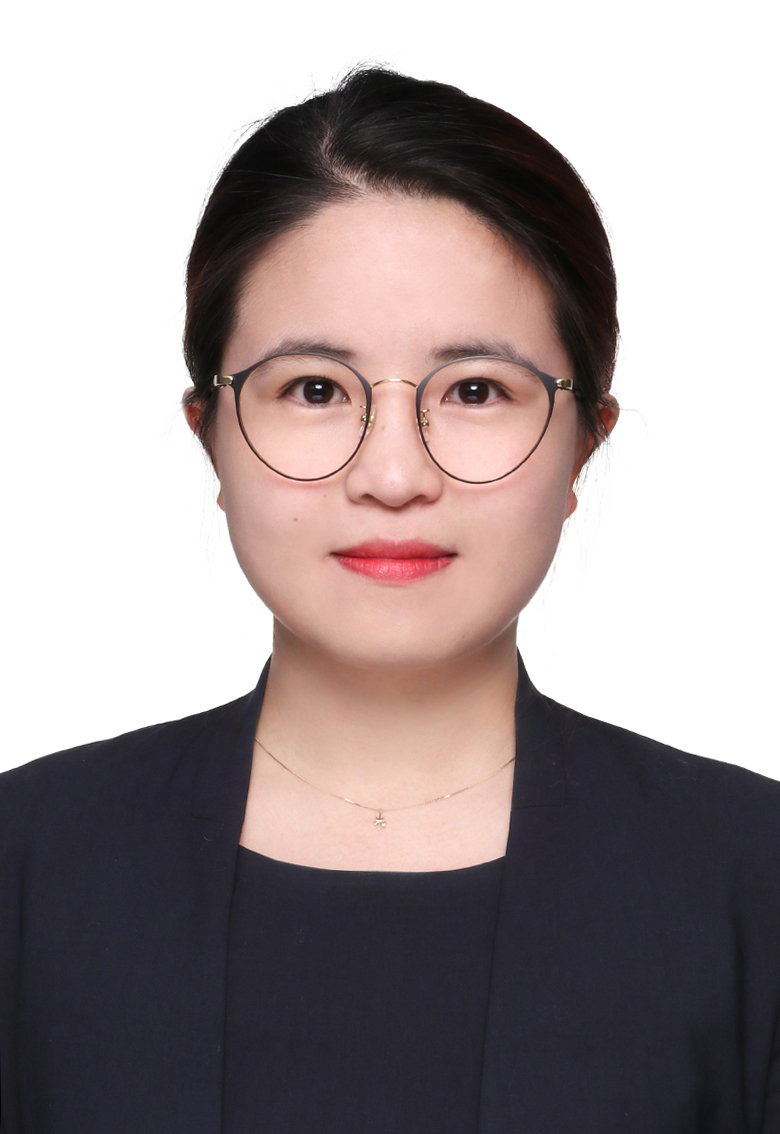}}]{Yuzhu Cui} was born in Xiangyang, Hubei, China, in 1991. She received the Ph.D. degree in Physics from National Astronomical Observatory of Japan and The Graduate University for Advanced Studies, Japan, in 2021.

 She is currently the Postdoc of Tsung-Dao Lee Institute with Shanghai Jiao Tong University. Her research interests include network science and big data analysis.
 \end{IEEEbiography}

\EOD

\end{document}